\documentclass[useAMS,usenatbib,usegraphicx]{mn2e}
\usepackage{amssymb}
\usepackage{amsmath}
\title[Mass segregation in hierarchical clusters]{How fast is
  mass-segregation happening in hierarchical formed embedded star
  clusters?}  

\author[R. Dom\'inguez et al.]{
  R. Dom\'inguez$^{1}$\thanks{E-mail:rdominguez@astro-udec.cl}, 
  M. Fellhauer$^{1}$, M. Bla\~na$^{2}$, J.P.~Farias$^{3}$,
  J. Dabringhausen$^{4}$ \\  
  $^1$Departamento de Astronom\'{i}a, Universidad de Concepci\'{o}n,
  Casilla 160-C, Concepci\'{o}n, Chile\\  
  $^2$Max-Planck-Institut f\"{u}r extraterrestrische Physik,
  Gie$\beta$enbachstra$\beta$e 1, 85748 Garching bei M\"{u}nchen,
  Germany\\ 
  $^3$Department of Astronomy, University of Florida, Gainesville, FL
  32611, USA\\
  $^{4}$Astronomical Institute, Charles University of Prague, V
  Holesovickach 2, 180 00 Praha 8, Czech Republic
}
\begin{document}

\label{firstpage}

\date{Accepted -----. Received -----; in original form -----}

\pagerange{\pageref{firstpage}--\pageref{lastpage}} \pubyear{2017}

\maketitle
  
\begin{abstract}
  We investigate the evolution of mass segregation in initially
  sub-structured young embedded star clusters with two different
  background potentials mimicking the gas.  Our clusters are initially
  in virial or sub-virial global states and have different initial
  distributions for the most massive stars: randomly placed, initially
  mass segregated or even inverse segregation.  By means of N-body
  simulation we follow their evolution for $5$~Myr.  We measure the
  mass segregation using the minimum spanning tree method
  $\Lambda_{\rm MSR}$ and an equivalent restricted method.
  Despite this variety of different initial conditions, we find that
  our stellar distributions almost always settle very fast into a mass
  segregated and more spherical configuration, suggesting that once we
  see a spherical or nearly spherical embedded star cluster, we can be
  sure it is mass segregated no matter what the real initial
  conditions were.  We, furthermore, report under which circumstances
  this process can be more rapid or delayed, respectively.
\end{abstract}


\section{Introduction}
\label{sec:intro}

Mass segregation (MS) has been detected in embedded star clusters
\citep{Lad96, Hil97, Hil&Har98, Bon&Bic06, Chen07, Er13}, i.e., that the
most massive stars are more centrally concentrated.  Advances in
observations have shown that star clusters form sub-structured
\citep{Lar95, Elm00, Tes00, Wil00, Car&Wit04, Gut05, Sch&Kle06,
  Car&Hod08, Sch08} and this has also been supported by simulations
\citep{Kle&Bur00, Kle&Bur01, Bat03, Bon03, Bat09, Off09, Gir12, Dal12}.
Observations also show that stars form dynamically cool (sub-virial
state) \citep{Per06, Per07, Pro09} and this property is supported by
hydrodynamical simulations of star formation \citep{Kle&Bur00, Off09,
  Mas10}.  A sub-virial initial state quickly erases primordial
substructure, producing a smooth and more spherical cluster
\citep{All09}.   

The process to form a segregated cluster can be dynamical
\citep{Mcm07, All09, Yu11} and fast for cool and sub-structured
clusters \citep{All10, Par16} reaching some level of MS in $\sim 1$
Myr.  Primordial MS is also possible, as a result of the star
formation process \citep{Zin82, Mur96, Elm01, Kle01, Bon01a,
  Bon&Bat06} where the massive stars tend to form in the centre of the
star-forming regions because of higher accretion rates in the central
parts or as a result of competitive accretion \citep{Lar82, Mur96,
  Bon97}.  The later has been used to explain very high levels of MS
in very young objects, when dynamical processes are not fast enough to
explain the observed MS \citep{Bon&Dav98, Rab&Mer98}.  On the other
hand \citet{Par15} pointed out that primordial mass segregation rarely
occurs if massive stars form by competitive accretion.  

In this paper, we show the evolution of MS for young embedded star
clusters, which form hierarchical out of fractal initial
distributions \citep[similar to][]{Par14}, but under the influence of
two different background (BG) potentials and their importance
depending on the locations of the massive stars.  We start our
simulations after the end of the star formation process where the
dynamics of the stars become important.  

In Section~2 we present the method and the different initial
conditions that we explore.  We continue in Section~3 with the
presentation of the results of our $N$-body simulations.  In Section~4 
we discuss our results focusing on the different evolution of MS
depending on the initial conditions.  We conclude in Section~5. 

\section{Method and Initial Conditions}
\label{sec:setup}

\subsection{Initial conditions of the stars}
\label{sec:fractal}

The stars are initially placed in a fractal distribution.

We use the method of \citet{Goo&Whit04} to generate initially
sub-structured distributions.  This method defines a cube of size
$N_{\rm div} = 2$ in which the fractal is created.  This cube is the
first-generation parent which is divided in $N^3_{\rm div}$ sub-cubes
(in our case $N^3_{\rm div} = 8$).  Each sub-node, called child, can
turn into a parent for a next generation with a probability of
$N^{D-3}_{\rm div}$ where $D$ is the fractal dimension, for which we
choose a $D = 1.6$.  The probability to become a parent is clearly
ruled by the value of the fractal dimension.  For lower values of $D$
less children turn into parents and so a more sub-structured
distribution is produced.  On the other hand, a value of $D = 3$
corresponds to an uniform distribution.  The not surviving children
are removed, then a small noise (displacement from the centre of the
cell of a few per cent of the cell-size) is added to the remaining
survivals to prevent an artificial grid-like structure and finally
they become parents for the next step.  The mature children are
divided into $N^{3}_{\rm div}$ new children of which again, using a
probability of $N^{(D-3)}_{\rm div}$, become a parent.  The process is
repeated until the number of children reaches a larger number than the
number of particles.  

The velocities of the parents follow a Gaussian with a mean of zero
and the children inherit the velocity from the parent including a
random component that decreases with each new generation.  That way
stars in the same area or clump have similar velocities and are not
flying in completely different directions.  The velocities get later
scaled to obtain the desired virial ratio (see further below).

We change the random seed value to produce different fractal
distributions with a maximum radius of $1.5$~pc.  

To the distribution of positions and velocities from the previous step
we assign masses using the modified initial mass function of
\citet{Kro02}, following:  
\begin{eqnarray}
    N(M) & \varpropto &
    \begin{cases}
      M^{-1.30} & \forall \ m_0 \leq M/{\rm M}_\odot < m_1  \\
      M^{-2.30} & \forall \ m_1 \leq M/{\rm M}_\odot < m_2  \\
      M^{-2.35} & \forall \ m_2 \leq M/{\rm M}_\odot < m_3
    \end{cases}
\end{eqnarray}
with $m_0=0.08$, $m_1=0.5$, $m_2=1.0$, $m_3=50$ M$_\odot$.  The
distribution is modified in the sense that we avoid the sub-stellar
mass range (below $0.08$~M$_\odot$ for brown dwarfs).  The total mass
of stars is $\sim 500$ M$_\odot$ for $\sim 1000$ stars which are
common values for these kind of objects \citep{Pis08}. 

\begin{figure}
  \includegraphics[scale=0.65]{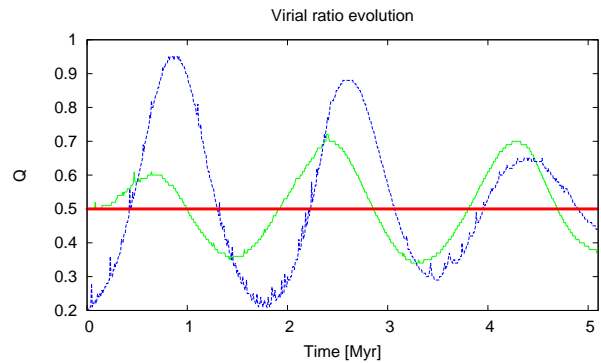},
  \caption{Example of the evolution of the virial ratio for a fractal
    distribution.  The green solid line shows the oscillations for a
    fractal starting from a pseudo virial equilibrium ($Q_{\rm init} =
    0.5$) and the blue dashed line shows the changes for the same
    fractal now starting from a cool state ($Q_{\rm init} = 0.2$).
    Horizontal line denotes the virial equilibrium value of $Q = 0.5$}  
  \label{fig:Qevol}
\end{figure}

\begin{figure*}
  \includegraphics[scale=0.42]{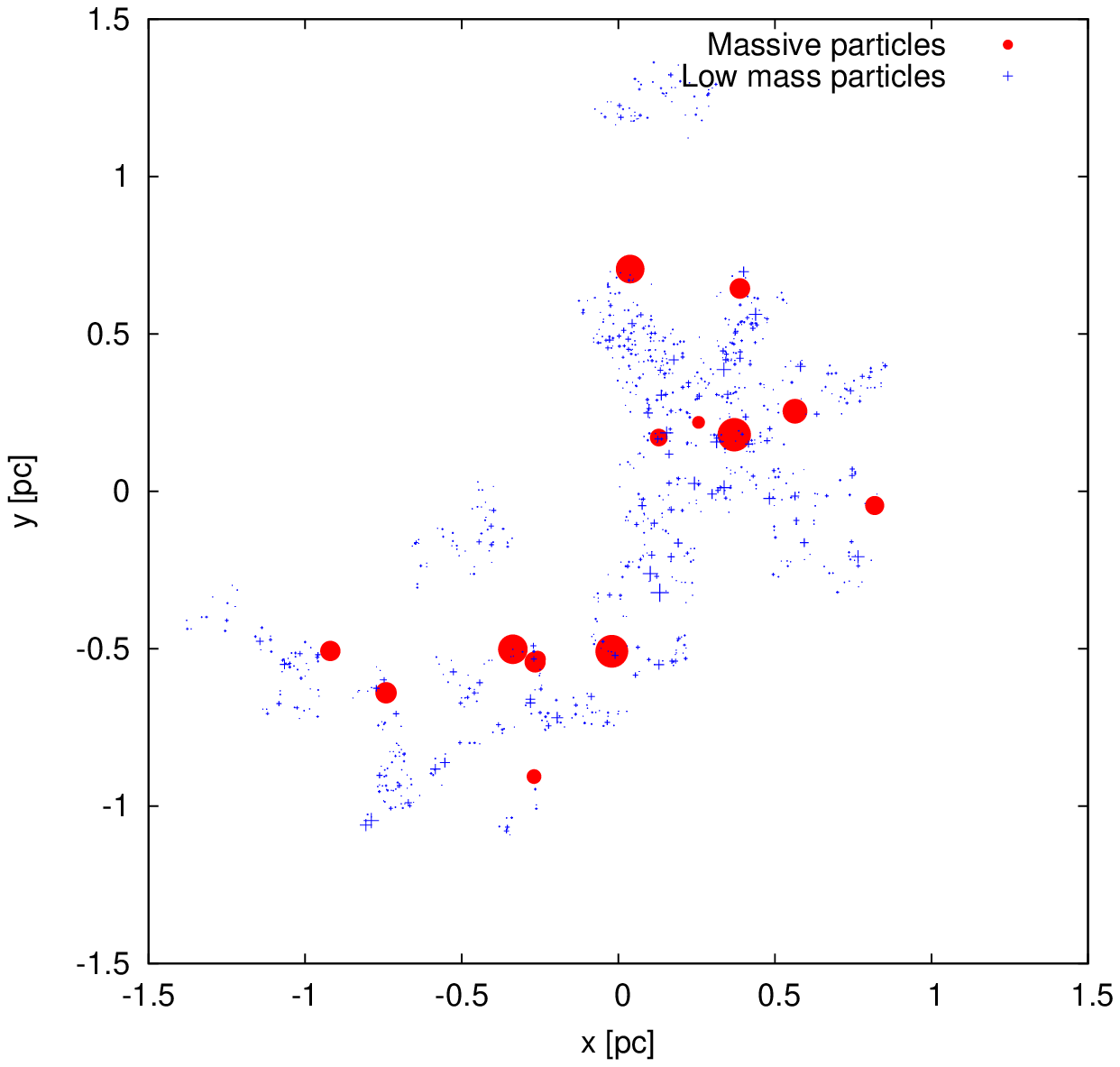},
  \includegraphics[scale=0.42]{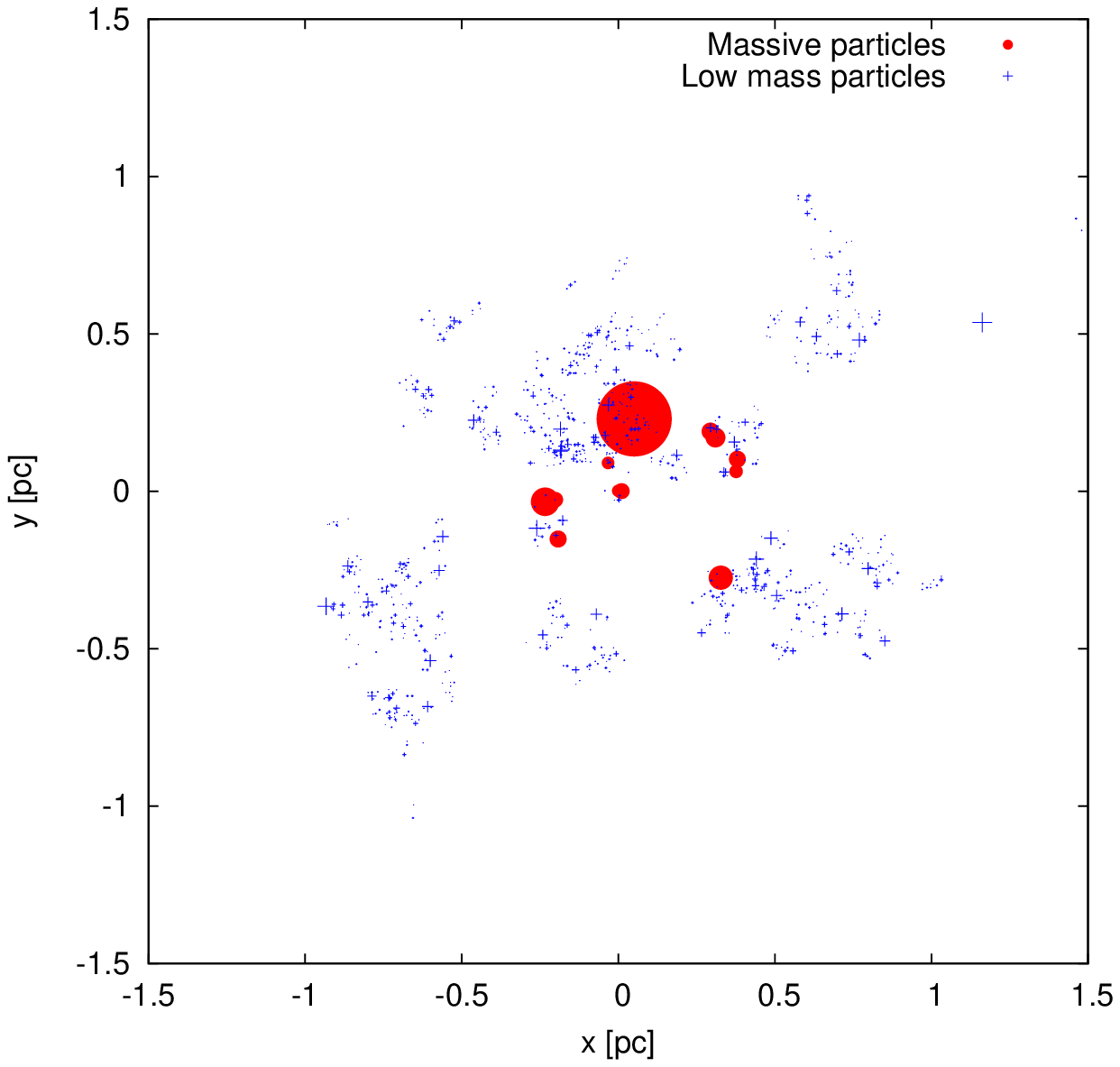},
  \includegraphics[scale=0.42]{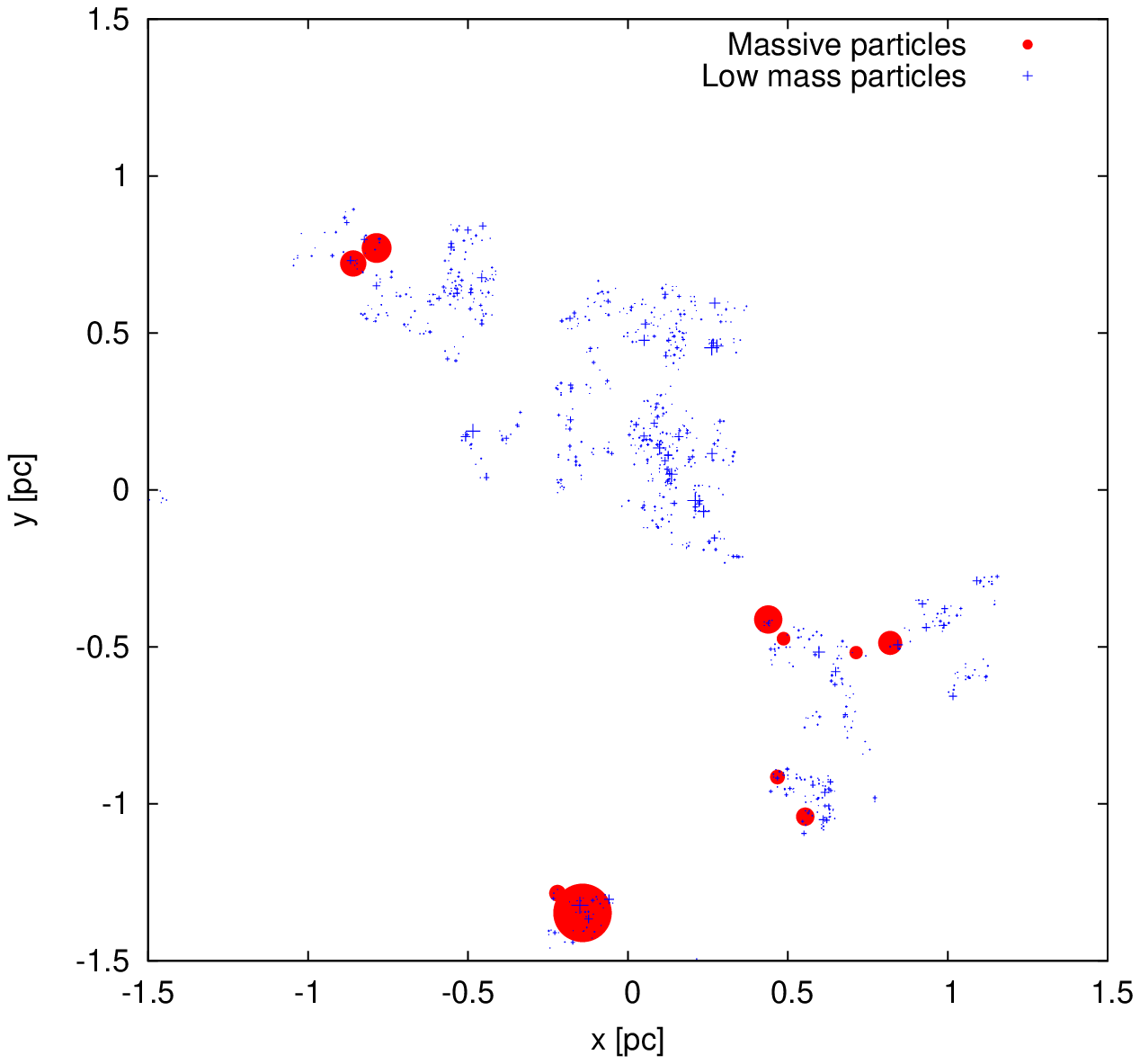}  
  \caption{Examples of the three different initial distributions of
    massive stars assumed in this work (shown in 2D).  Red filled
    circles are high-mass stars ($M \geq 4$~M$_{\odot}$) and blue points
    are low-mass stars ($M < 4$~M$_{\odot}$). 
    The sizes of circles and points are proportional to the mass of
    the stars.  Left panel: initial fractal with random placement of
    the masses (NO-SEG).  Central panel: initial fractal with
    inside-out segregation (SEG-IN), i.e.\ the highest mass stars are
    in the centre of the distribution.  Right panel: initial fractal
    with outside-in segregation (SEG-OUT), i.e.\ the highest mass are
    furthest away from the centre.} 
  \label{fig:massdist}  
\end{figure*}

In our simulations, we investigate the influence of the initial virial
ratio as well.  The virial ratio is defined as:
\begin{eqnarray}
  \label{eq:Q}
  Q & = & \frac{T}{|\Omega|},
\end{eqnarray}
i.e.\ the ratio between kinetic ($T$) and potential ($\Omega$) energy.
We start our simulations with values of $Q_{\rm init} = 0.2$ and
$Q_{\rm init} = 0.5$.  The values we achieve by re-scaling the kinetic
energy, changing the velocities of all stars in a proportional manner.  

A value of $Q = 0.5$ usually signifies that the system is in
virial equilibrium, i.e.\ it is stable and should not be subject to
changes.  This is only true for well populated spherical systems and
not for the sub-structured distributions, we are using.  

Our stellar distributions try to re-arrange themselves into a smaller
in size but close to spherical distribution.  This produces an initial
compression, reaching higher values of $Q$ and then oscillating
around the virial state until a stable spherical shape is produced. 
Starting with a value of $Q_{\rm init} = 0.2$ will produce a strong
compression, reaching even higher values of $Q$ than in the
previous case and then oscillating around the virial state as well,
erasing the fractal distribution even faster.  An example of this
process is shown in Fig.~\ref{fig:Qevol}.  The green dashed line shows
the evolution of a fractal with $Q_{\rm init} = 0.5$ and the blue
solid line for the same fractal now starting with $Q_{\rm init} = 0.2$. 

We use three different sets of initial distributions for the massive
stars.  As a massive star we define all stars with $M \geq
4$~M$_\odot$.  Stars less massive are regarded as low-mass stars in
our investigation.

The different distribution can be described as follows:
\begin{itemize}
\item In the first set we place the masses randomly throughout the
  fractal distributions, and we refer to this configuration as not
  segregated: NO-SEG.  
\item In the second set we force all massive stars to be located
  within $0.5$~pc and we call it initially mass segregated: SEG-IN.   
\item In the third set, we locate all massive stars at radii larger
  than $1.0$~pc to mimic inverse mass-segregation: SEG-OUT.  
\end{itemize}

The SEG-OUT case is very unlikely but we analyze it to test the
accepted trend of MS occurring no matter the initial conditions.  In
Fig.~\ref{fig:massdist} we show the projected distribution of one
representative fractal for each initial massive star distribution.
Red filled circles are high-mass stars and blue points are low-mass
stars.  The different sizes of circles and points are proportional to
the mass of the stars.  The left panel shows an initial fractal with
NO-SEG placements of the masses.  The central panel is an initial
fractal with SEG-IN configuration.  The right panel shows the initial
fractal with SEG-OUT placement.  The number of high-mass stars is
ruled by the random sampling of the IMF, but we ensure that in
each random realization at least eight massive stars are present. 

\subsection{Mimicking the gas distribution}
\label{sec:back}

To explore the impact of the distribution of gas, we mimic the
remaining gas by using two different analytic background (BG) potentials.

For the first set of simulations we use a Plummer potential
\citep{Plu11}:
\begin{eqnarray}
  \label{eq:plummer}
  \rho(r) & = & \frac{3{M}_{\rm Pl}} {4\pi R_{\rm Pl}^3} \left( 1 +
    \frac{r^2}{R^2_{\rm Pl}}\right)^{-\frac{5}{2}} \nonumber \\
  M(r) & = & M_{\rm Pl}\frac{r^3}{R^3_{\rm Pl}}\left( 1 +
    \frac{r^2} {R^2_{\rm Pl}}\right)^{-\frac{3}{2}} \\
  \Phi(r) & = & - \frac{G M_{\rm Pl}}{R_{\rm Pl}} \left( 1 +
    \frac{r^2}{R^2_{\rm Pl}}\right)^{-\frac{1}{2}} \nonumber
\end{eqnarray}
with ${M}_{\rm Pl}$ and $R_{\rm Pl}$ the Plummer Mass and Plummer
radius respectively.

The second set of simulations follows a uniform density profile of the
form:  
\begin{eqnarray}
  \label{eq:uniform}
  \rho(r) & = & \frac{3M_{\rm tot}} {4\pi r^3_{\rm c}} \nonumber \\
  M(r) & = & M_{\rm tot} \frac{r^{3}}{r^3_{\rm c}} \\
  \Phi(r) & = & \frac{G M_{\rm tot}} {2 r^{3}_{\rm c}} \left( r^{2} -
    3 r_{\rm c}^{2} \right) \nonumber
\end{eqnarray}
with ${M}_{\rm tot}$ the total mass of the sphere.  The radius of the
uniform sphere is given by $r_c$. 

We choose $M_{\rm Pl} = 3472$~M$_{\odot}$ and $R_{\rm Pl} = 1.0$~pc.
For uniform case we have $M_{\rm tot} = 3455$~M$_{\odot}$ and $r_{\rm
  c} = 1.8$~pc.  Both distributions ensure that we have exactly a
gas-mass of $M_{\rm gas}(1.5 {\rm pc}) = 2000$~M$_{\odot}$ within the
stellar distribution, which extends out to a radius of $1.5$~pc.  As
we have $\sim 500$ M$_\odot$ in stars, this ensures a star formation
efficiency (SFE) of $\epsilon = 0.2$. 

\subsection{The $\Lambda_{\rm MSR}$ parameter}
\label{sec:mst}

There are many methods to measure MS in a cluster, for example the
minimum spanning tree method $\Lambda_{\rm MSR}$ \citep{All09}, the
$\sum_{\rm LDR}$ method \citep{Mas&Cla11} and a technique where every
distance of the most massive stars to the centre of each group is
measured \citep{Kir&Mye11,Kir14}.  

Some methods were tested by \citet{Par&God15} resulting, in many
cases, in contradictory results as they define MS differently.  The
authors concluded that the $\Lambda_{\rm MSR}$ method is able to measure
MS not only in the classical cases but it is also very efficient and
produces reliable results for sub-structured regions.   

Based on this study we use the $\Lambda_{\rm MSR}$ method in our
investigation.  One advantage in using this method, especially for
clusters with substructures, is that it is not necessary to define a
centre, where the density is highest.  In a sub-structured
distribution, one might not have a clear high density centre or even
worse two or more over-densities within the cluster.  Another
advantage is that this method includes an associated error, giving
information of how reliable the measurement of the MS is.  

This method measures the length of the minimum spanning tree (MST;
shortest connection between all stars of the sample without
crossings) of the $N$ most massive stars $l_{\rm massive}$ and
compares it with the MST of the same number of low-mass stars.  As we
have many combination possibilities as there are many more low-mass
stars present, we choose in total $1600$ different random samples of
low-mass stars and calculate a mean length of their MSTs $\langle
l_{\rm norm}\rangle$ as well as the respective standard deviation
$\sigma_{\rm norm}$: 
\begin{eqnarray}
  \Lambda_{\rm MSR} & = & \frac{\langle l_{\rm norm} \rangle} {l_{\rm
      massive}} \pm \frac{\sigma_{\rm norm}} {l_{\rm massive}}  
  \label{eq:Lambda}
\end{eqnarray}
where $\Lambda_{\rm MSR}$ is the degree of MS.  $\Lambda_{\rm MSR}
\sim 1$ indicates no MS, $\Lambda_{\rm MSR} >> 1$ indicates strong MS
and $\Lambda_{\rm MSR} < 1$ means inverse MS.   

In our study we use $N = 8$, as this is the minimum number of high
mass stars in our samples of the IMF.

\begin{figure}
  \includegraphics[scale=0.6]{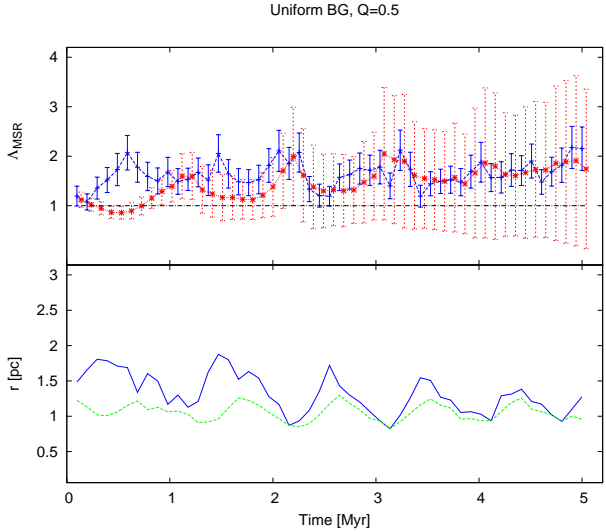}
  \caption{Example of one fractal with an initial mass distribution
    SEG-OUT, $Q_{\rm init} = 0.5$ and an uniform BG.  Top panel shows
    the evolution of $\Lambda_{\rm MSR}$ as red dashed line and stars.
    The evolution of the $\Lambda_{\rm rest}$ method is denoted with
    the blue solid line and plus signs.  For clarity the $\Lambda_{\rm
      MSR}$ symbols are slightly shifted in time.  The bottom panel
    compares the evolution of $r_8$ (blue solid line) and the
    value of $R_h$ (green dashed line).} 
  \label{fig:errorcomp}
\end{figure}

\subsection{Restricted minimum spanning tree method ($\Lambda_{\rm
    rest}$)}  
\label{sec:restmst}

The measurements of $\Lambda_{\rm MSR}$ can be very sensitive to low
mass stars getting kicked out due to strong two-body interactions
increasing the values of $l_{\rm norm}$.  This can be fixed to some
degree, repeating the measurements for $l_{\rm norm}$ many times (in
our case $1600$; see above).

A different source of error, can be the escape of one massive star
from the cluster, increasing the length of $l_{\rm massive}$
significantly.  A similar effect has a massive star, e.g.\ in the
SEG-OUT configuration, on a rather circular orbit around the cluster,
i.e.\ not taking part in any interactions and therefore not subject to
any process driving it towards the centre.  

Trying to avoid these problems, we restrict the measurements of
$\Lambda_{\rm MSR}$ by three conditions: 
\begin{enumerate}
\item For the determination of $l_{\rm massive}$ we use all
  massive stars inside of the half-mass radius ($R_{\rm h}$). 
\item If it is not possible to find at least eight of the massive
  stars inside of $R_{\rm h}$, we increase the search radius until we
  find the eighth massive star $r_{8}$ and base the determination
  of MS on this radius.
\item For the determination of $l_{\rm norm}$ we only use low-mass
  stars inside of $2r_8$, which can be equal to $2R_{\rm h}$, when at
  least eight of the massive stars are inside of $R_{\rm h}$. 
\end{enumerate} 
This restricted method we call $\Lambda_{\rm rest}$.  We
compare the values obtained from this restricted method with the
unrestricted ones (see Sect.~\ref{sec:mst}) in all our simulations.

In Fig.~\ref{fig:errorcomp}, we show the evolution for a fractal with
initial mass distribution SEG-OUT as an example.  In the top panel 
we show the evolution of $\Lambda_{\rm MSR}$ in red stars connected by
a dashed line and $\Lambda_{\rm rest}$ as blue plus-signs
connected by a solid line.  In the bottom panel we show the evolution
of $R_{\rm h}$ (green dashed line) and $r_{8}$ (blue solid line) for
the same simulation.

First, we note that towards the end of the simulation both methods
give within the errors the same final result which is somewhat
reassuring.  At the start of the simulation we see significant
differences between the two methods.  The unrestricted method shows a
very low value of MS, as expected for an inside-out simulation,
reaching a value of two (a significant value for MS) after more than
$2$~Myr and leveling out at a final value of about two.  In contrary the
restricted method shows a level of MS which is increasing immediately,
reaching a value above two in about half a Myr.  Furthermore, the
unrestricted method shows much larger error-bars.  

How can we understand this behaviour?  This simulation has more than
8 massive stars ($12$).  If some stars are not taking part in MS
(as explained above) or get ejected, the range of values for
$\Lambda_{\rm MSR}$ can be large leading to a large dispersion of the
results reflected in the error-bars.  It also gives us the expected low
values to begin with, as we are starting with the high-mass stars in
the outskirts of the cluster.  The restricted method focuses on the 
innermost $N = 8$ stars and shows that they are moving almost
immediately to the centre.  Both methods show strong oscillations of
the $\Lambda$ values, leveling off after about $3$~Myr, showing within
the errors slightly higher values for the restricted method than for
the unrestricted. 

What is the 'true' value of MS?  We would argue, there is no 'correct'
value for the MS, as it depends on your point of view.  If you focus
on the inner parts of the newly formed cluster only, you will detect a
sufficient level of MS almost immediately after just half a Myr (as
shown with the restricted results).  But, if you have the whole region,
in which the stars have formed, in sight, then you detect ejected
stars and stars orbiting the new cluster at larger distances, giving
you a whole range of MS values, depending on the sample of high-mass
stars you pick (large error-bars at the end of the simulation for the
unrestricted method).  This ambiguity is also visible in the
difference between $r_{8}$ and $R_{\rm h}$.  In the SEG-OUT
simulations we see that $r_{8}$ is always larger than $R_{\rm h}$ and
only at later stages they start to agree.

\begin{figure*}
  \includegraphics[scale=0.63]{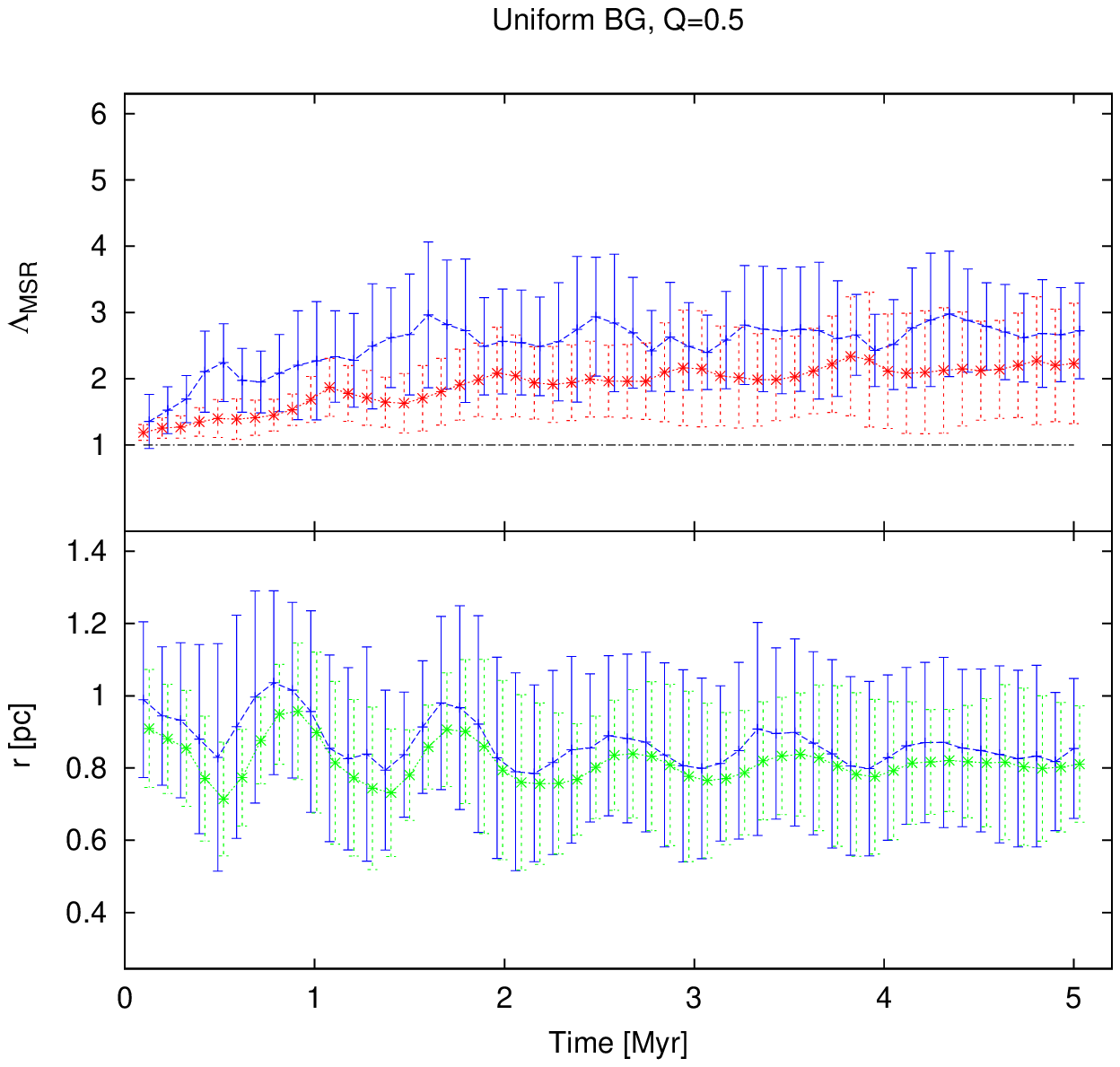} 
  \includegraphics[scale=0.63]{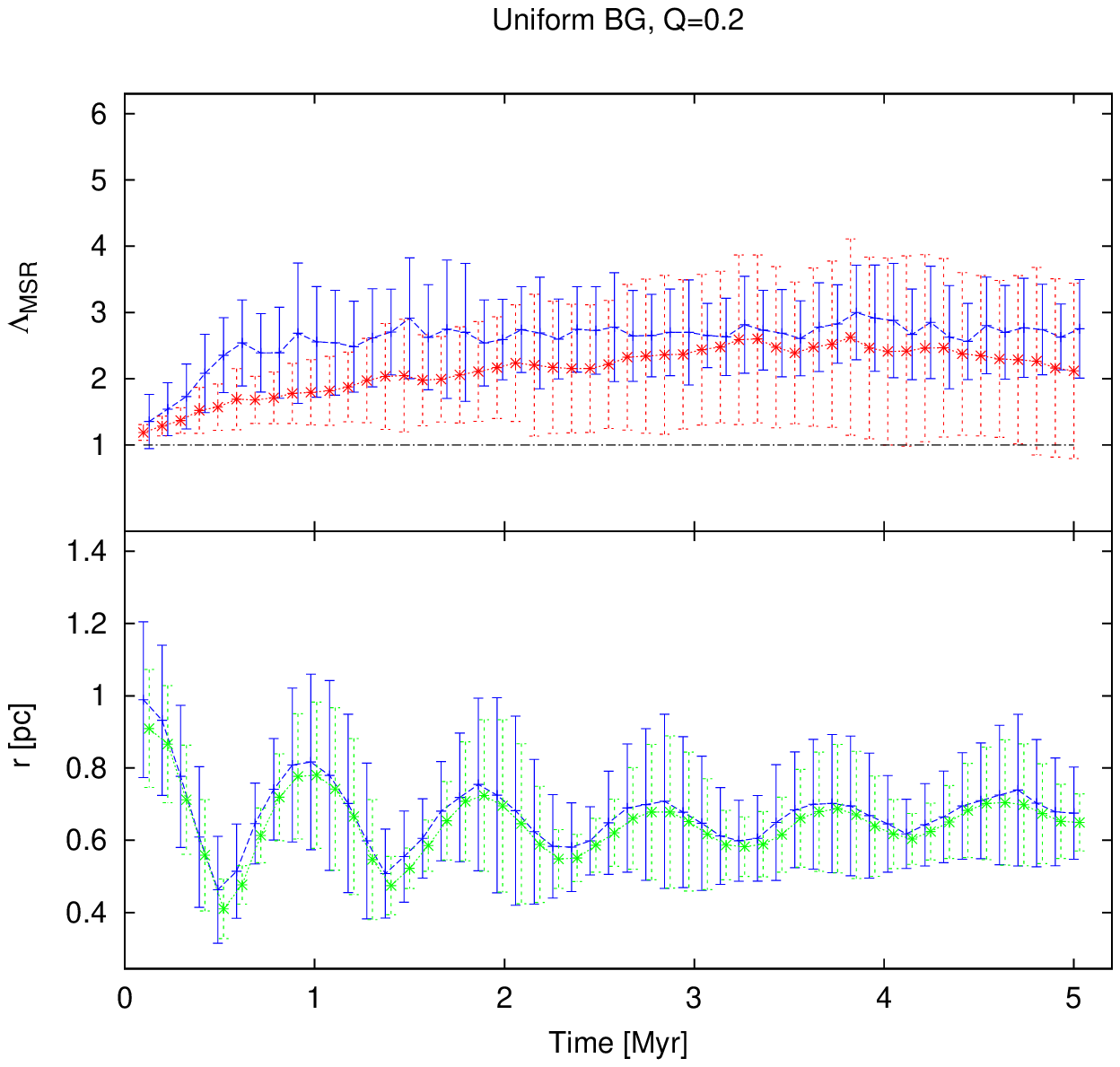}
  \includegraphics[scale=0.63]{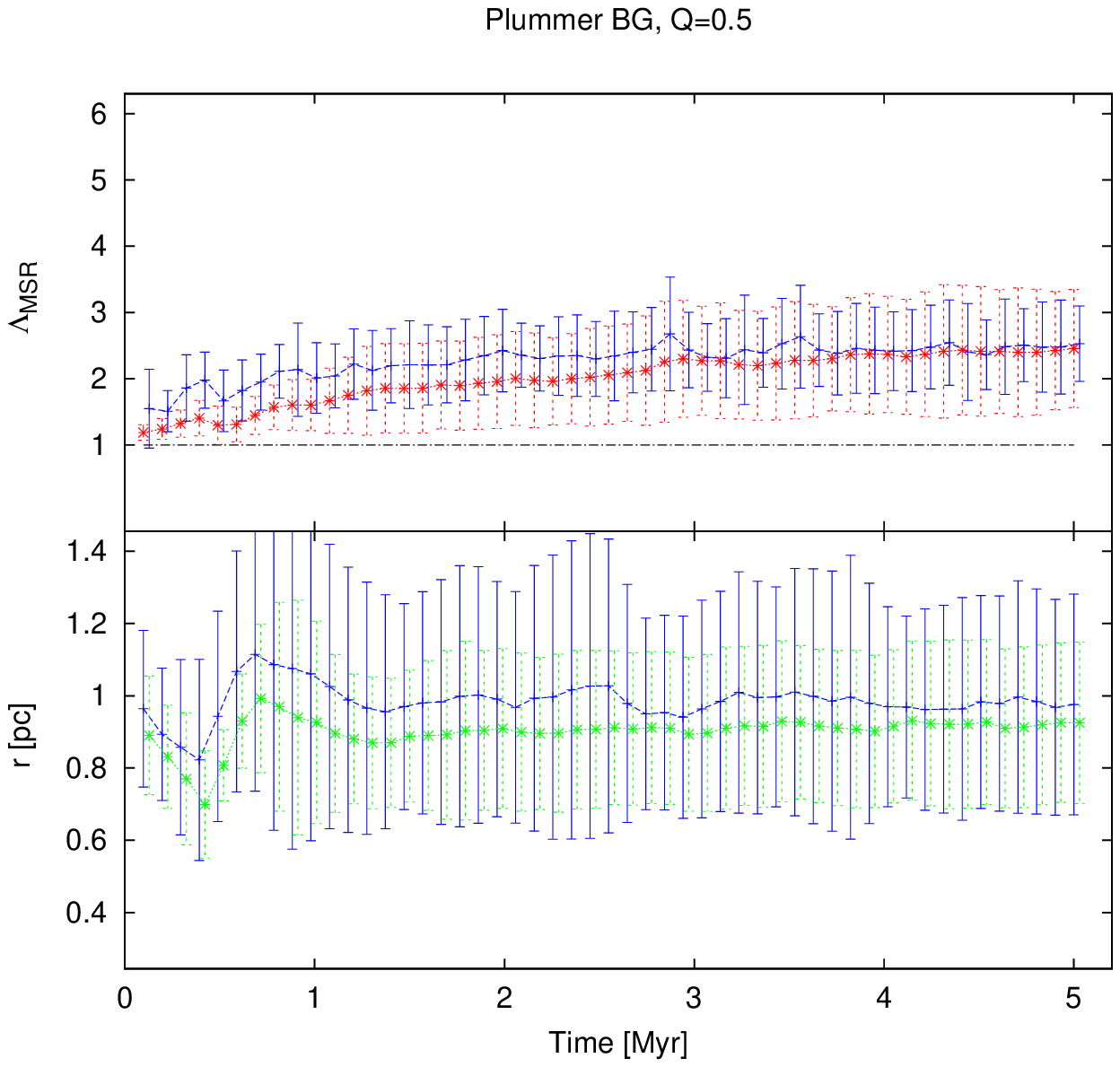} 
  \includegraphics[scale=0.63]{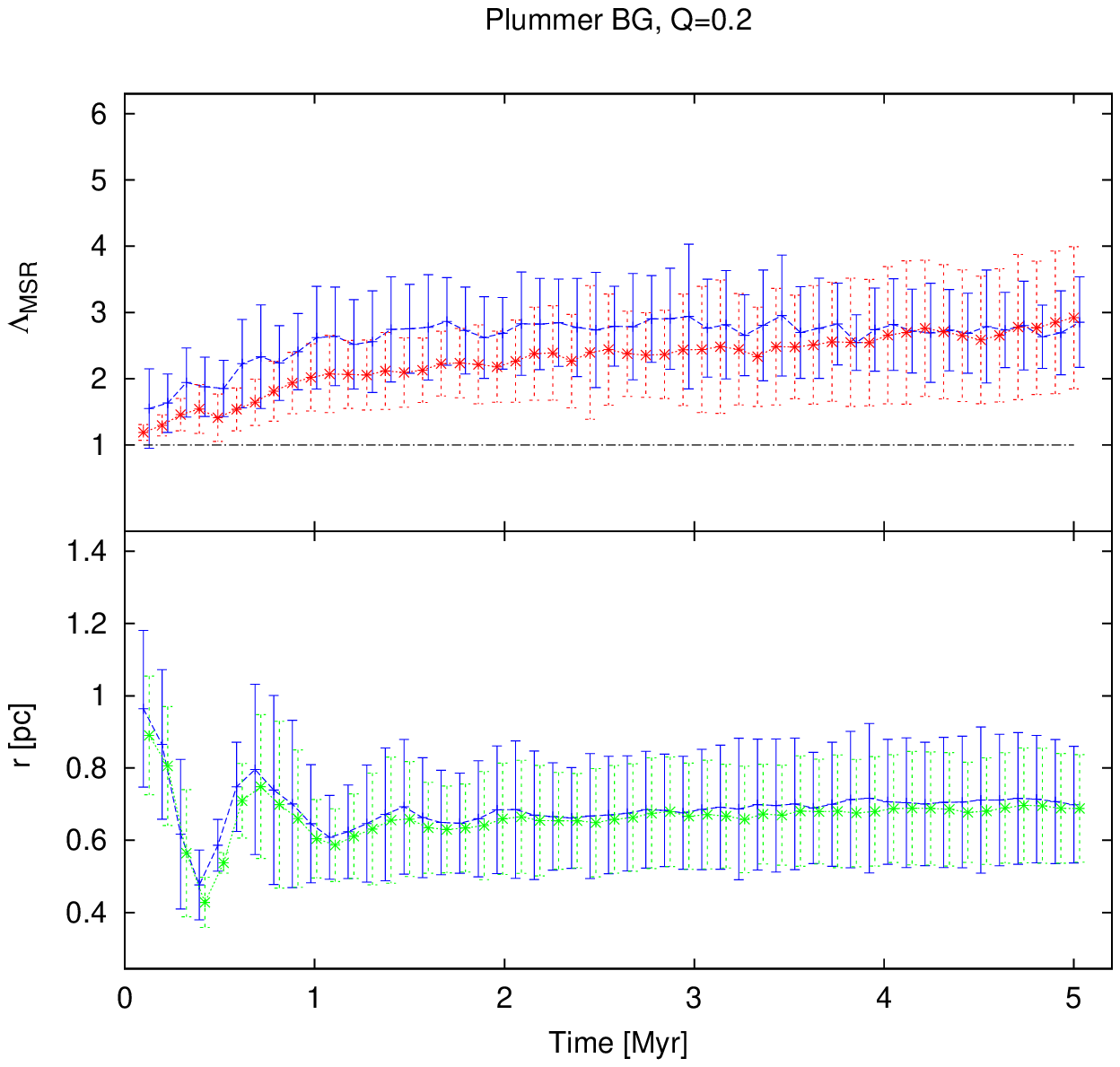}
  \caption{Results for NO-SEG, i.e.\ the masses are distributed
    randomly throughout the whole fractal.  In contrary to
    Fig.\ref{fig:errorcomp} each data-point is a mean value calculated
    from ten different random realisations.  Top row panels show the
    results for a uniform background and the lower row panels for the
    Plummer background.  Left panels are the simulations with $Q_{\rm
      init} = 0.5$, i.e.\ starting in 'pseudo-virial equilibrium' and
    right panels show the simulations where velocities are reduced to
    obtain $Q_{\rm init} = 0.2$.  In each panel the top half shows the
    evolution of $\Lambda_{\rm MSR}$ (red stars and dashed lines)
    and $\Lambda_{\rm rest}$ (blue plus signs and solid lines).  In
    the lower halves of the panels we show the evolution of $R_{\rm
      h}$ (green dashed line) and $r_{8}$ (blue solid line).}
  \label{fig:PNOSEG}
\end{figure*}

\begin{figure*}
  \includegraphics[scale=0.63]{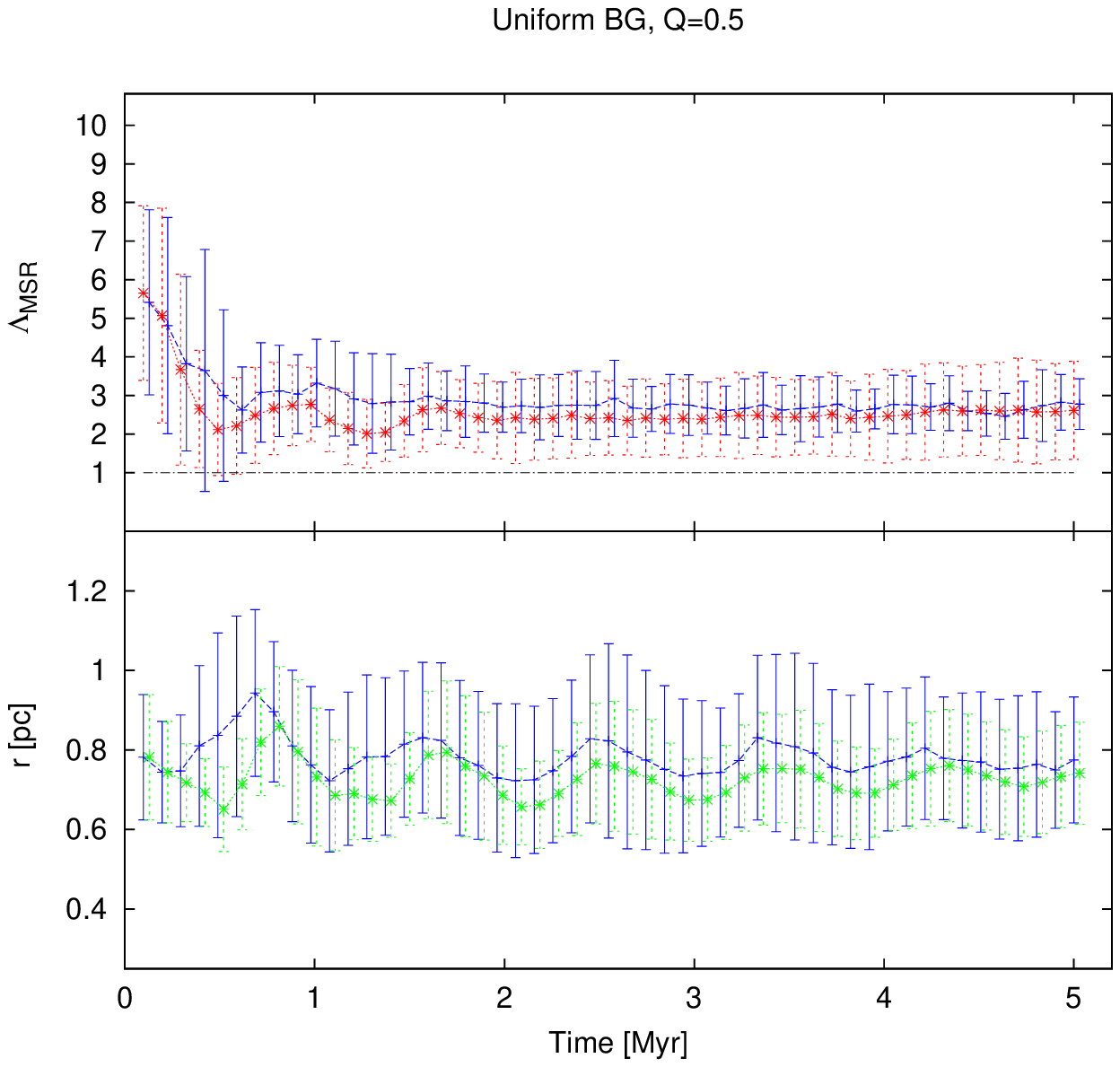} 
  \includegraphics[scale=0.63]{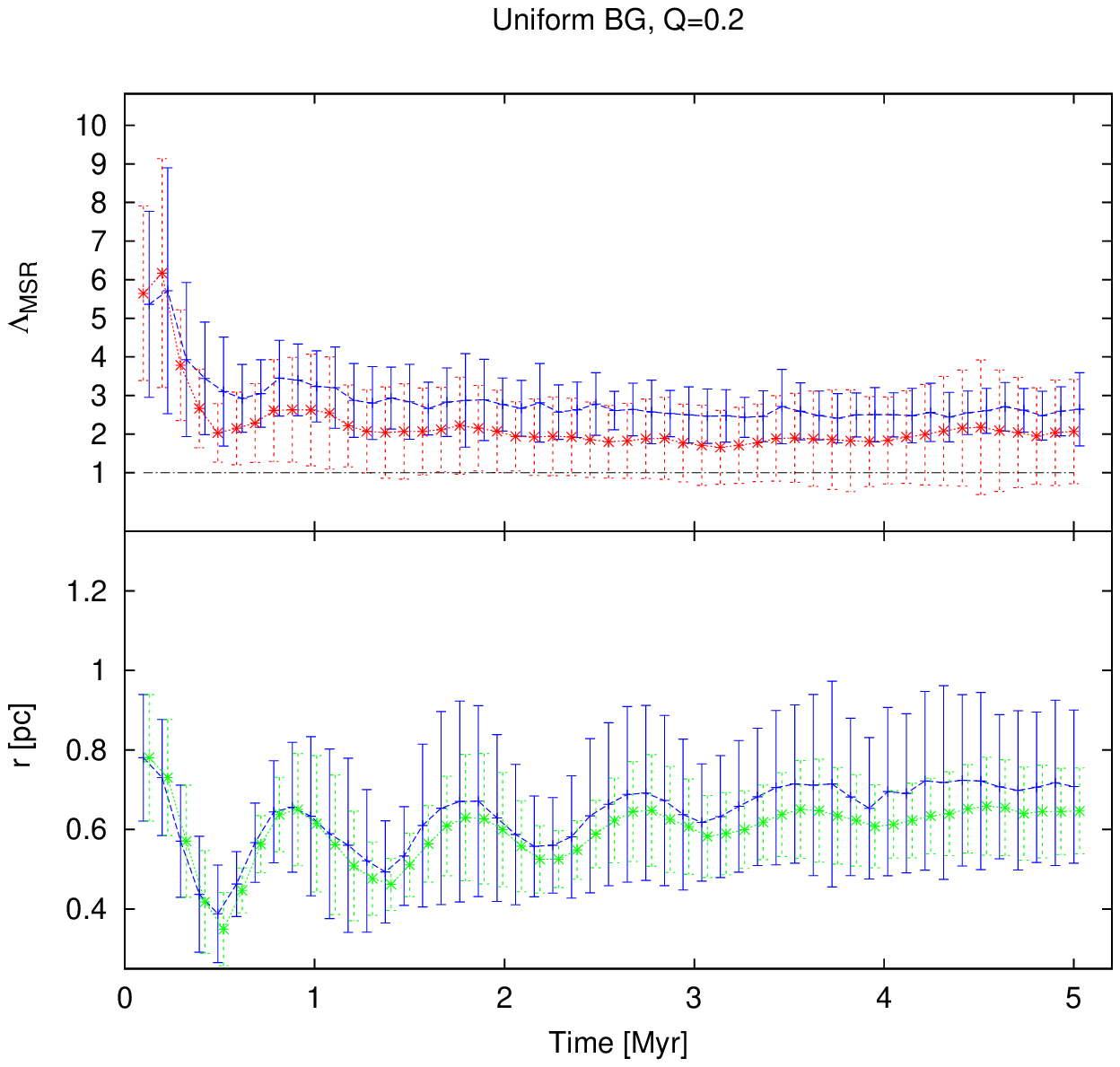}
  \includegraphics[scale=0.63]{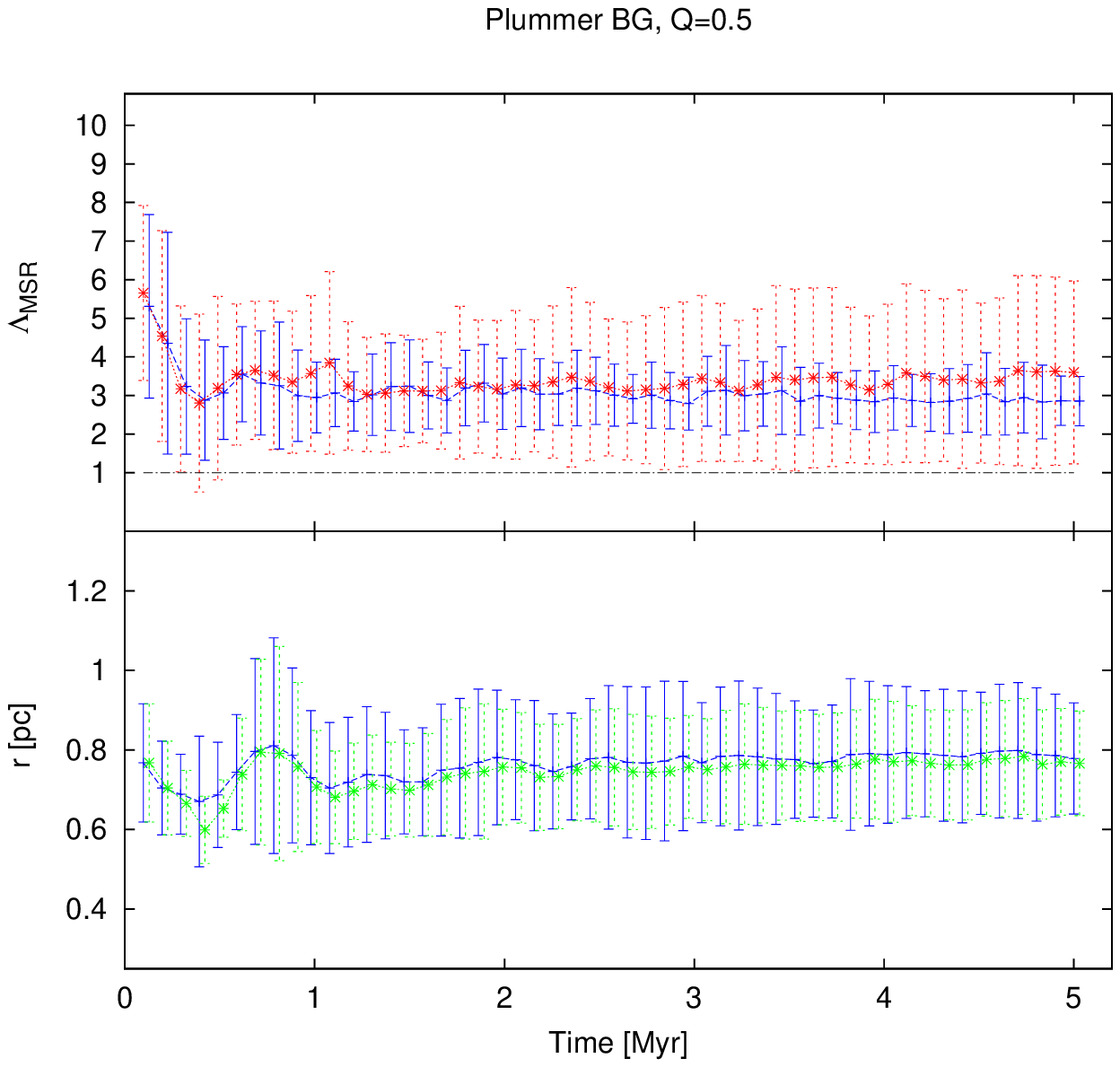}
  \includegraphics[scale=0.63]{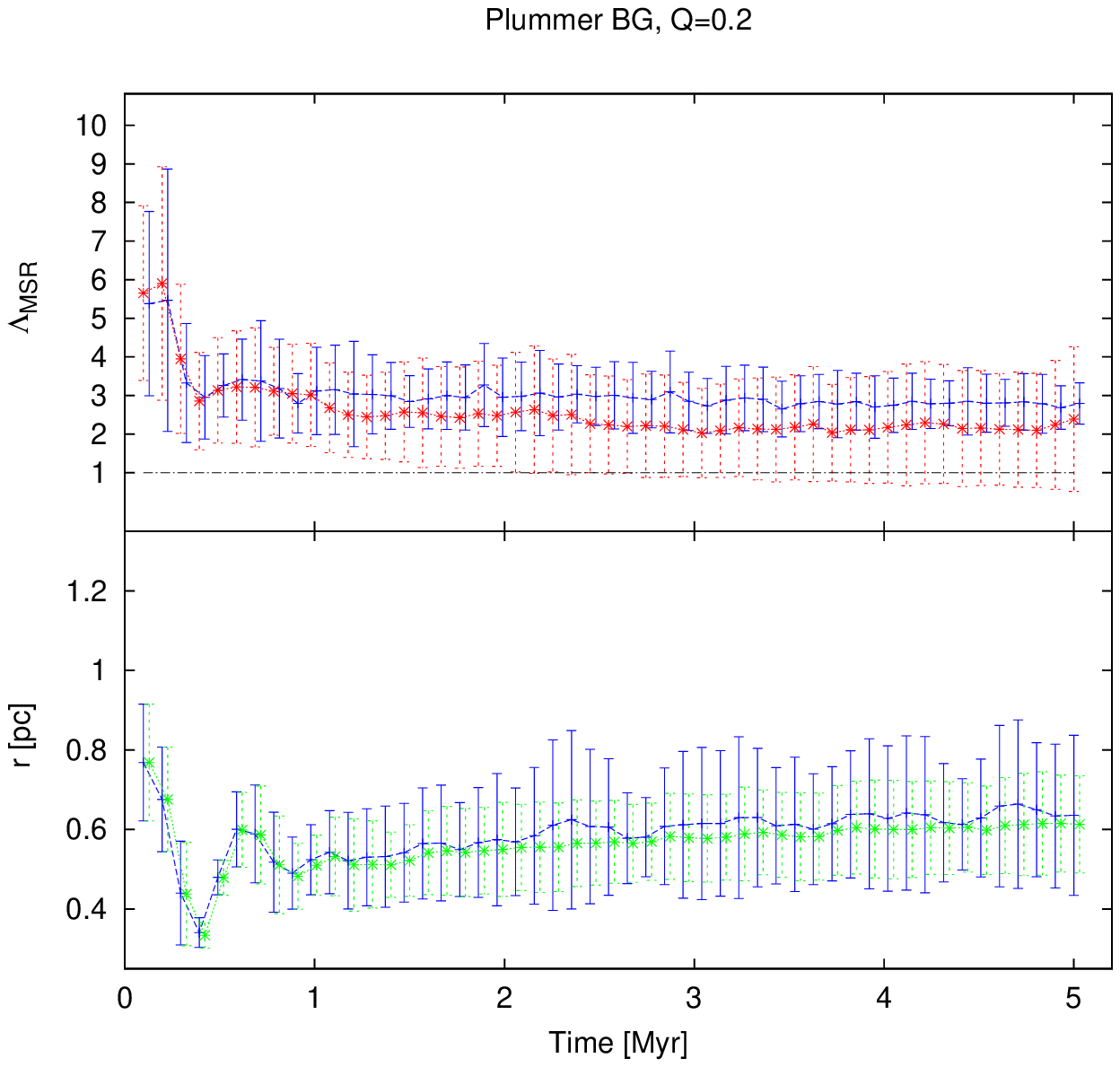}
  \caption{Same as in Fig.~\ref{fig:PNOSEG} but now for the SEG-IN
    simulations, i.e.\ where all the massive stars are initially
    within the central $0.5$~pc.}
  \label{fig:PSEGIN}
\end{figure*}

\begin{figure*}
  \includegraphics[scale=0.6]{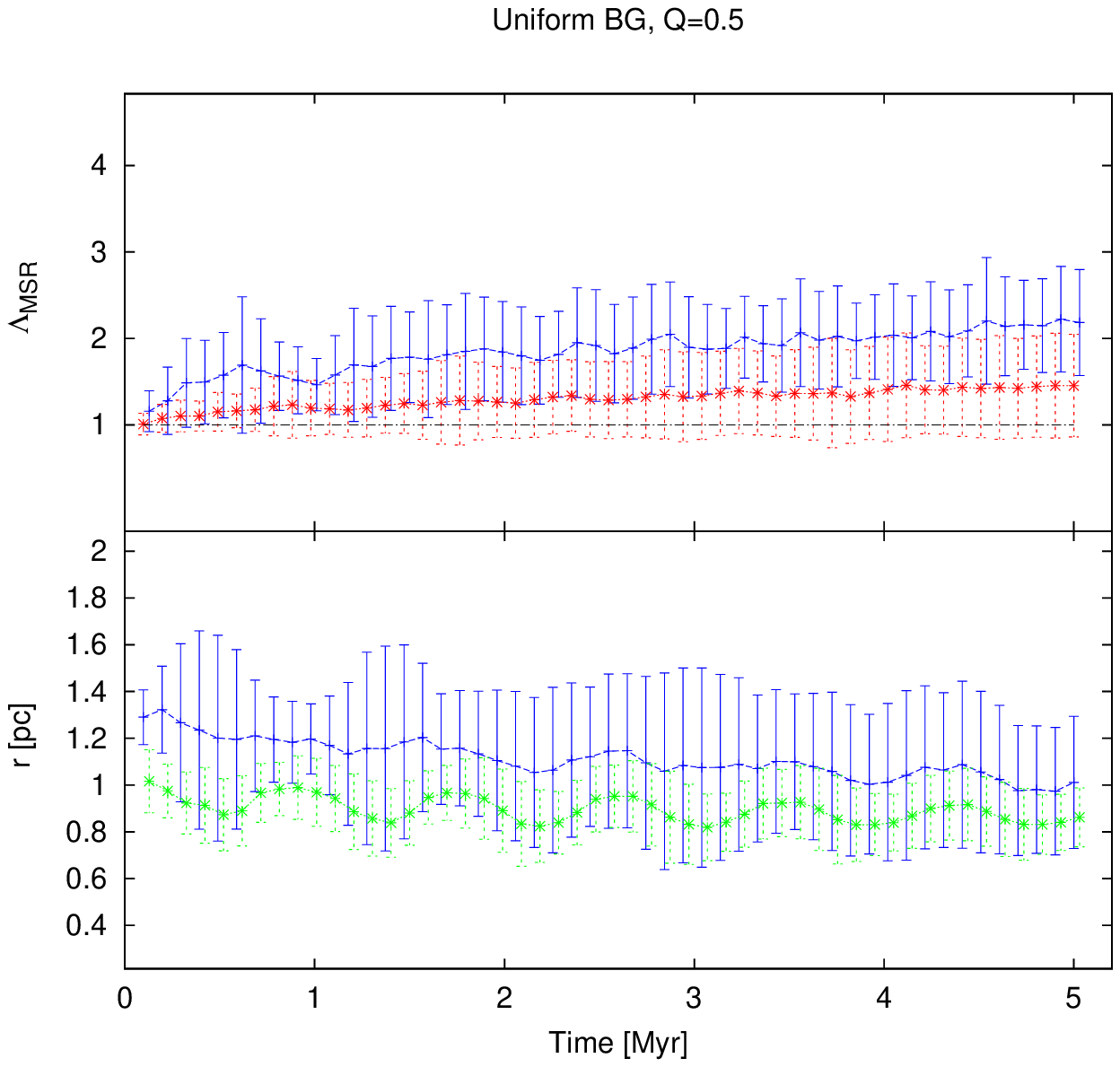} 
  \includegraphics[scale=0.6]{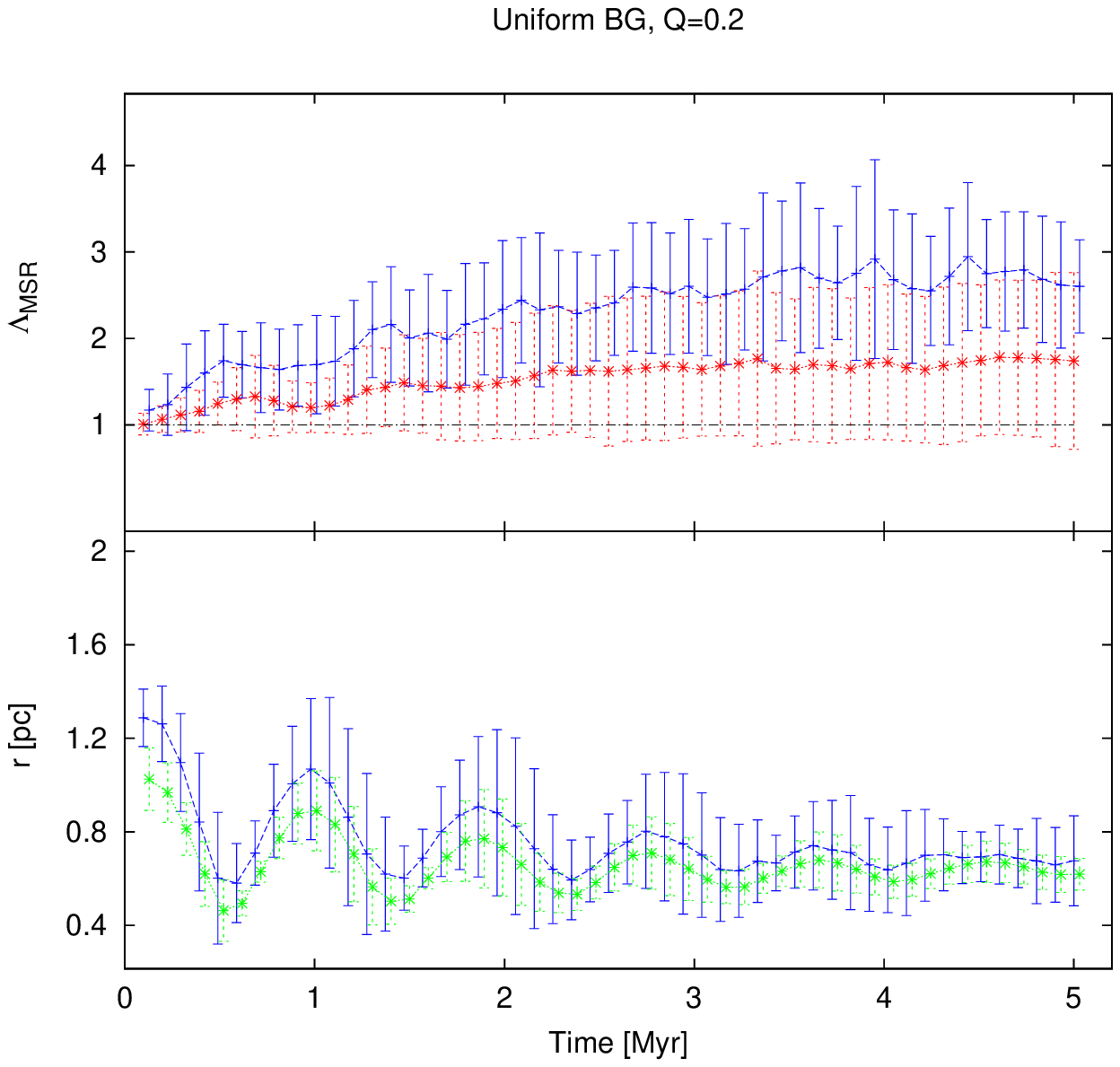}
  \includegraphics[scale=0.6]{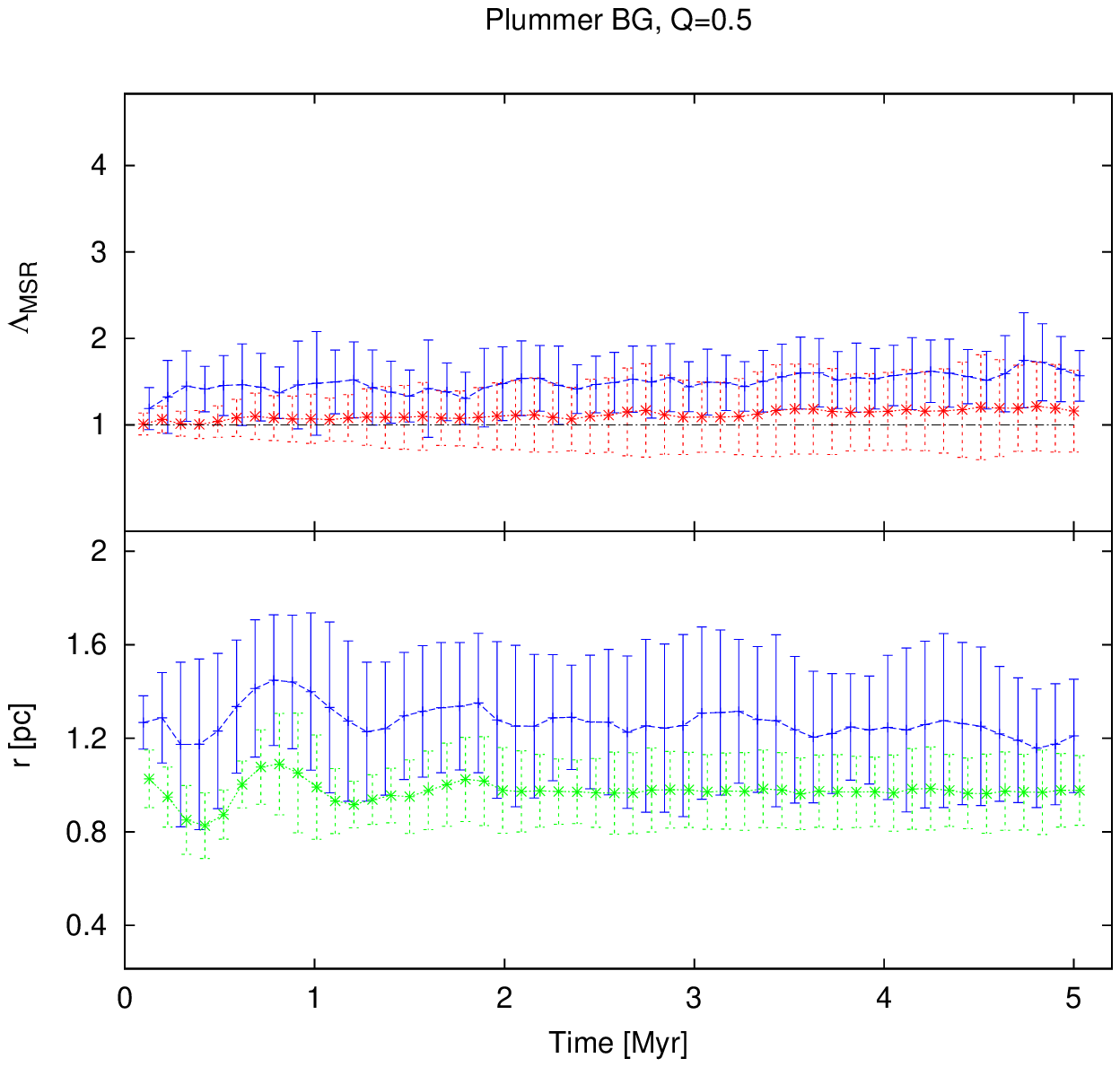} 
  \includegraphics[scale=0.6]{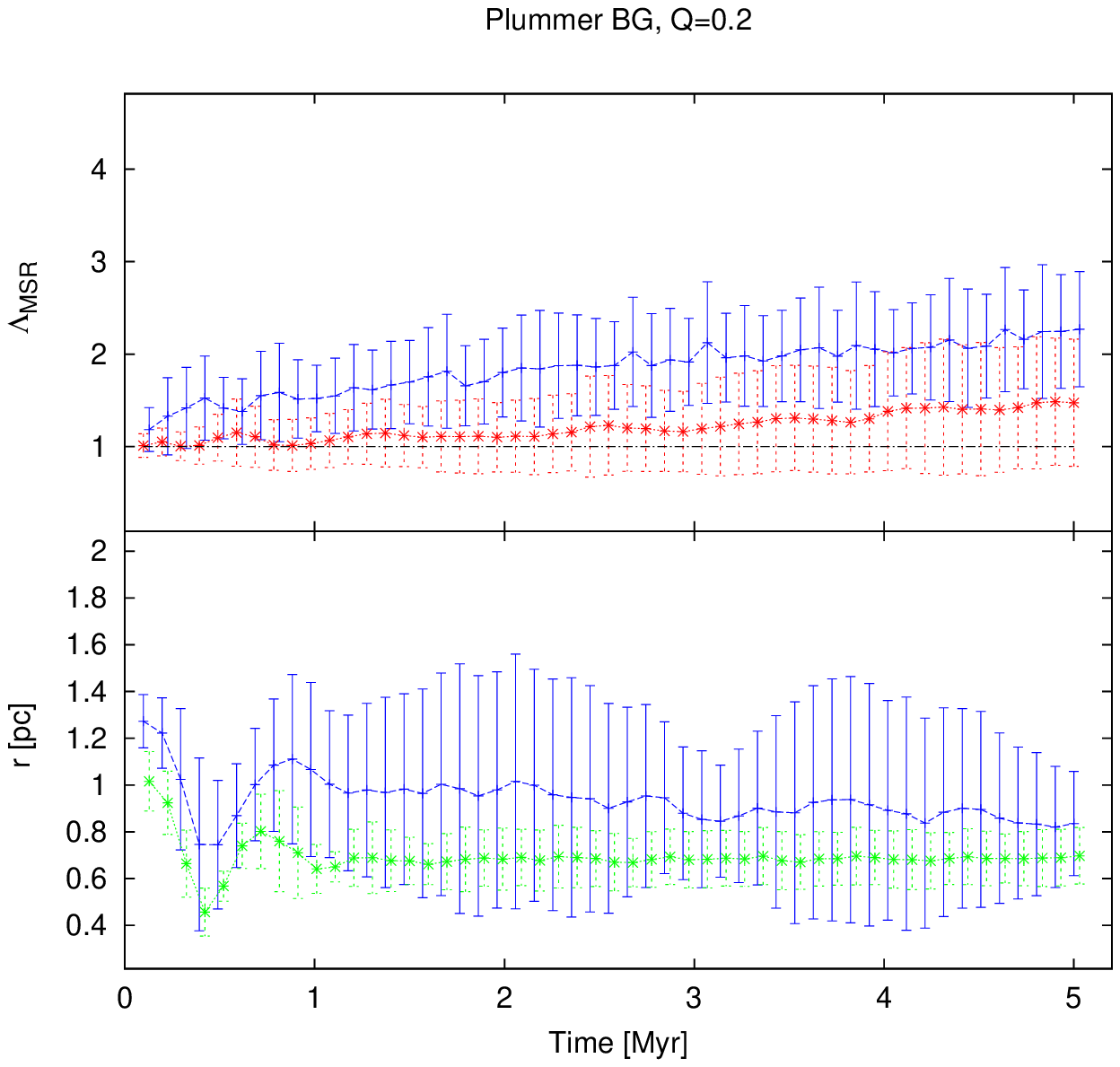}
  \caption{Same as in Fig.~\ref{fig:PNOSEG} but now for the SEG-OUT
    simulations, i.e.\ where all the massive stars are initially
    outside the central $1.0$~pc.}
  \label{fig:PSEGOUT}
\end{figure*}

\subsection{Set of simulations}
\label{sec:simset}

We perform, in total, $480$ N-body simulations using the direct N-body
integrator code NBODY~6 \citep{Aar03} with $1000$ different mass
particles with sub-virial, virial and clumpy initials conditions
embedded in two different BG potentials.  

We use $12$ different fractal distributions for the initial
positions.  For each fractal we draw a different IMF sample.  We use
four fractals with their respective IMF samples (position--mass pair)
for each set of initial mass placement cases (NO-SEG, SEG-IN and
SEG-OUT).  For each pair of positions and masses we generate 10
different random assignments of the masses to the positions, following
the general rule of each case.  This gives us for each case $40$
different sets of initial conditions ($120$ in total), thereby
overcoming the influence of by-chance results far from the mean values.

All these sets are now placed into two different background potentials
(uniform and Plummer) and are simulated using velocities scaled to a
pseudo equilibrium ($Q_{\rm init} = 0.5$) or to a cold case with $Q_{\rm
  init} = 0.2$.

\section{Results} 
\label{sec:results}

In the following sub-sections we present the results for the different
initial MS states, namely NO-SEG, SEG-IN and SEG-OUT.

The figure shown for each sub-section shows in the left panels the
pseudo-virial initial conditions and in the right panels the results
for the sub-virial initial state.  The top panels are the uniform
background simulations, while the lower panels show the results for
the background potential which follows a Plummer distribution.  The
upper half of each panel shows the evolution of the MS ($\Lambda_{\rm
  MSR}$ and $\Lambda_{\rm rest}$ respectively) and the lower half the
evolution of $R_{\rm h}$ and $r_{8}$.  In all figures we show the mean
values taken from all $40$ realisations of this set of parameters,
instead of the results of one single simulation.   Again, the
unrestricted method is shown as red stars and dashed lines, while the
restricted values are given as blue plus-signs and solid lines in the
upper part of each panel, while the lower parts show $r_{8}$ in solid
blue and $R_{h}$ in dashed green. 

\subsection{NO-SEG}
\label{sec:no-seg}

In Fig.~\ref{fig:PNOSEG}, we see in the top parts of each panel, that
both methods pick up a $\Lambda$ value (here the term $\Lambda$ refers
to both methods) close to one initially.  The restricted method shows
a slightly larger $\Lambda$ value than the unrestricted one, an efect
we have discussed above and we will come back to in the discussion
below.  The simulations end up with mean values between $2.0$ and
$3.0$.   

In the top left panel (uniform BG, $Q_{\rm init} = 0.5$) we see an
increase until about $1.5$~Myr.  Afterwards, we see small oscillations
around a value of slightly below $3$.  At about $4.3$~Myr the highest
degree of MS is found ($\Lambda_{\rm rest} = 2.98$).  As a final value
we assign a mean value of $\Lambda_{\rm rest}$ measured from all
data-points between $4$ and $5$~Myr called $\Lambda_{\rm final}$,
which in this case is $\Lambda_{\rm final} = 2.68 \pm 0.86$.  We call
the time where the level of $\Lambda_{\rm rest}$ remains larger than
$2$, (i.e. a significant level of MS) $t_{\rm seg}$.  In this set of
simulations we find $t_{\rm seg} = 0.39$~Myr.   

In the bottom half of this panel we see small oscillations in $R_{\rm
  h}$, which are damping out with time and settling at a slightly
lower value than the initial one, i.e.\ we see that even in the
supposedly virialised case we get a net contraction of the final
cluster. 

The top-right box is showing the results for the same background
potential but now starting in a cool state ($Q_{\rm init} = 0.2$),
i.e.\ undergoing a strong collapse at the beginning.  Starting with
the same initial degree of MS (as both sets rely on the same initial
fractals and mass assignments, just with differently scaled velocities)
a stronger increase of $\Lambda_{\rm rest}$ is observed reaching the
highest value of $\Lambda_{\rm rest} = 3.01$) at already $1.47$~Myr.
After that, the variations are smaller than in the 'virial' case
leading to a final mean value of $\Lambda_{\rm final} = 2.76 \pm
0.82$.  Again, both methods agree well within their errors, with the
restricted method showing slightly larger values of MS.  The value of
$t_{\rm seg} = 0.39$~Myr. 

In the bottom half of this panel we can see the strong decrease of
$R_{\rm h}$, due to the cold initial conditions.  Afterwards, we obtain
large oscillations until the simulations settle into a final value
much lower than in the pseudo-virialised case. 

The bottom-left panel shows the results for the Plummer BG potential
starting in pseudo-equilibrium.  We see a slow rise in MS leveling
off at a final value of $2.51 \pm 0.78$.  The highest value of $2.68
\pm 0.91$ is reached after $2.84$~Myr.  For $t_{\rm seg}$ we obtain
$0.78$~Myr.  Both methods give indistinguishable values towards the end
of the simulation.

\begin{figure*}
  \includegraphics[scale=0.4]{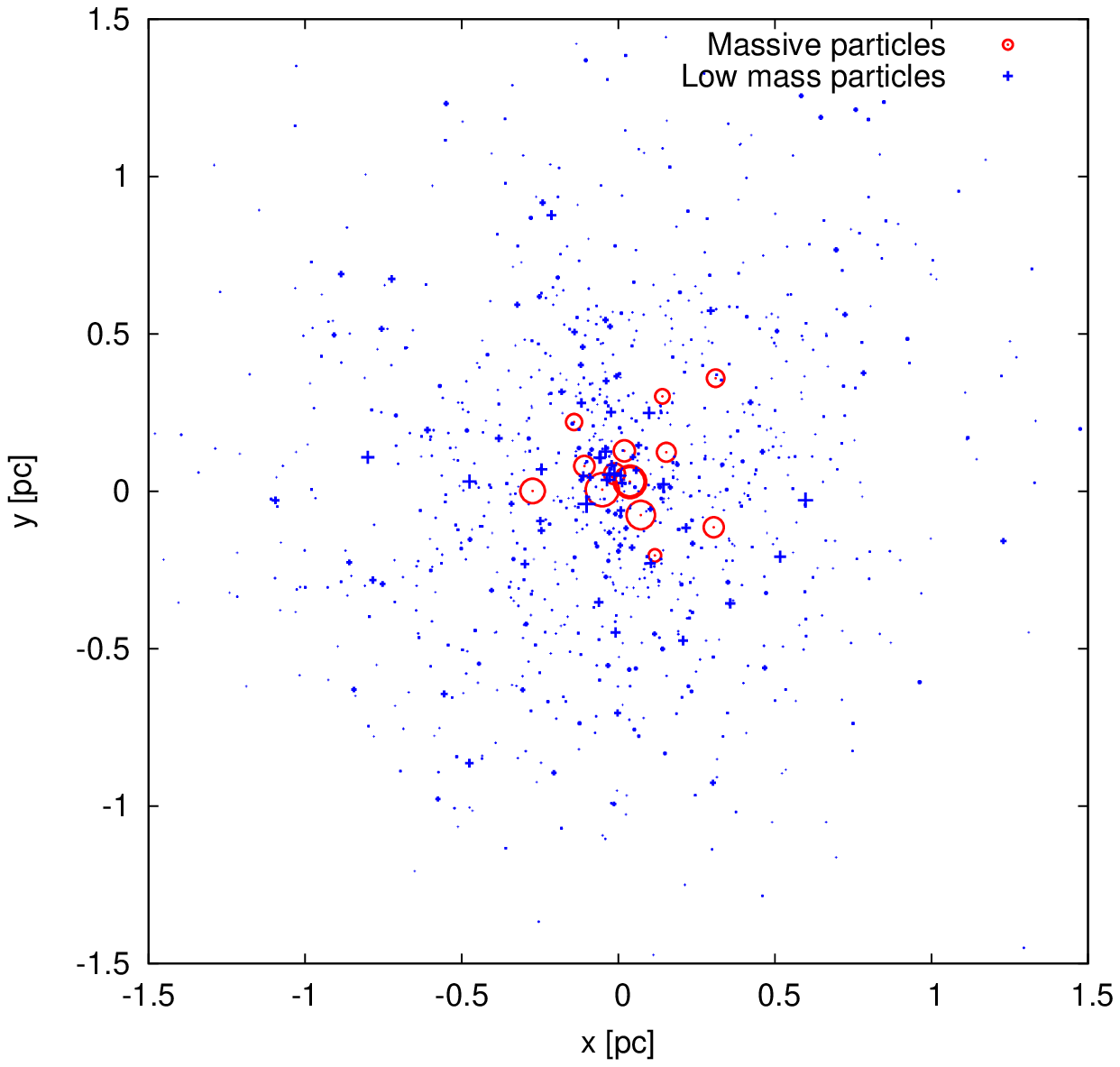},
  \includegraphics[scale=0.4]{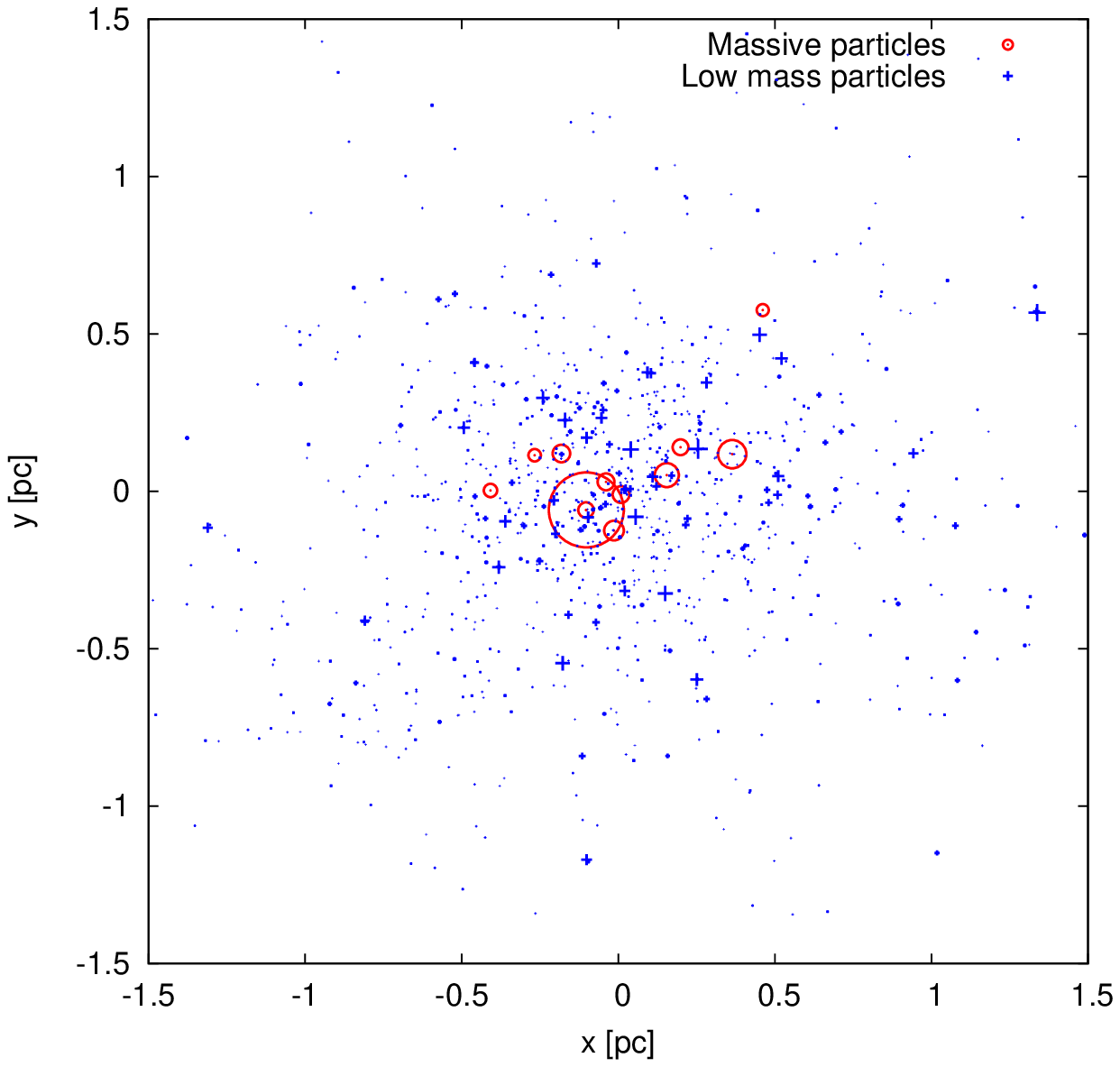},
  \includegraphics[scale=0.4]{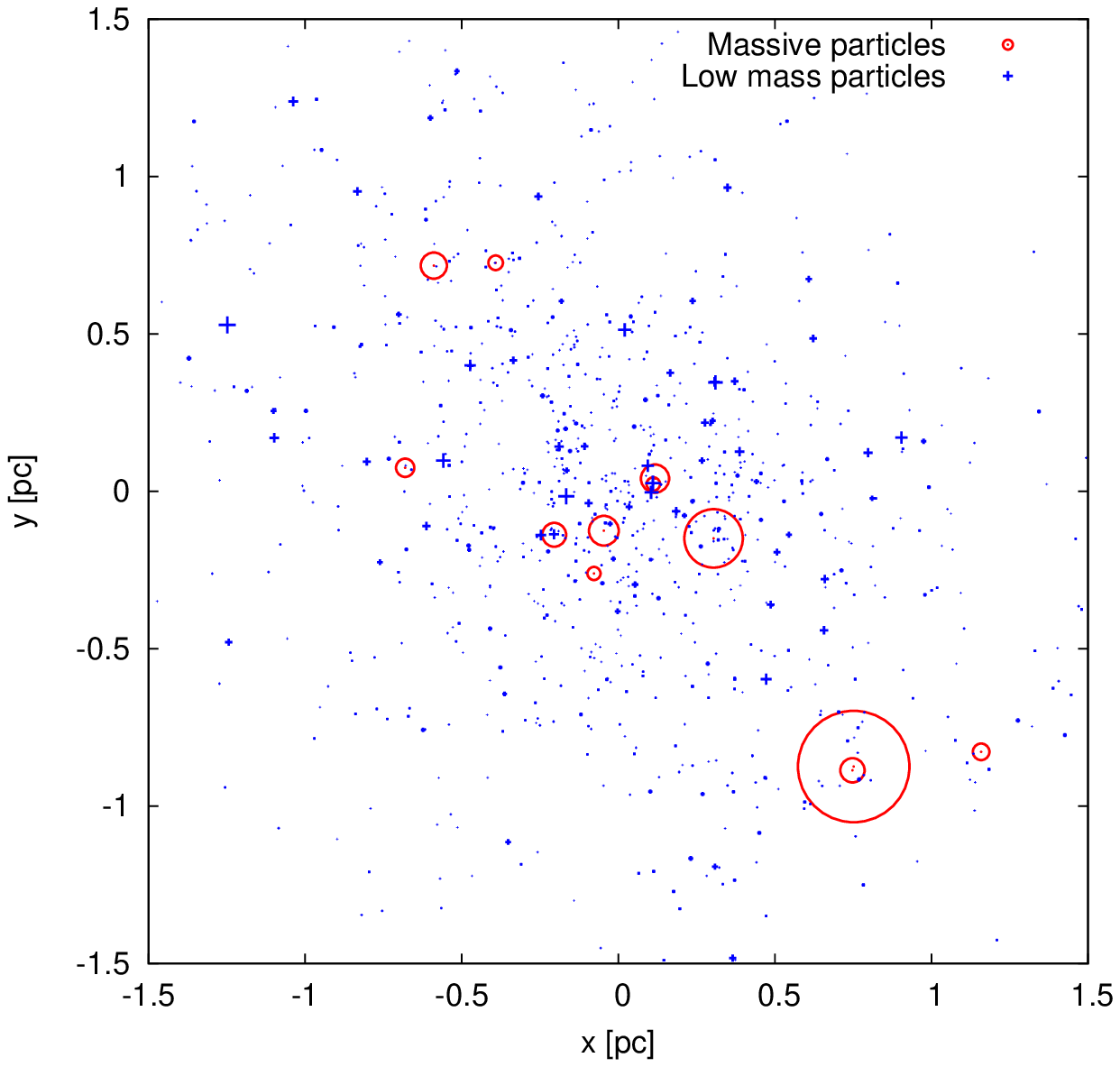}
  \caption{Examples of the final state for the three different initial
    distribution of massive stars shown in Fig.~\ref{fig:massdist}.
    Left panel: Final state of a fractal which starts with NO-SEG.
    Central panel: Final state of a fractal which starts with SEG-IN.
    Right panel: Final state of a fractal which starts with SEG-OUT.} 
\label{fig:massdistfin}
\end{figure*}

In the bottom half of this panel we see larger error-bars for $R_{\rm
  h}$.  This is explainable by the fact that in a uniform sphere we
have a well defined free-fall time which is the same for all radii and
so all particles more or less reach the densest configuration at the
same time, which is also more or less the same for all simulations of
the previous set of parameters.  Now, in the Plummer BG case, the
free-fall times vary with radius and the point of densest
configuration can vary in time for the different sets of initial
particle distributions.  Therefore, we see no clear oscillations in
our results, which are mean values calculated from all simulations,
but rather inflated error-bars. 

The bottom-right panel shows the results for the Plummer BG potential
starting in a cool state i.e.\ we get a strong compression.  We
observe a similar trend as in the cold uniform background case: a rise
in MS and then an almost constant plateau.  The maximum value of
$\Lambda_{\rm rest} = 3.03$ is reached after $2.16$~Myr.  The final
value $\Lambda_{\rm final} = 2.74$.  We measure a $t_{\rm seg} =
0.59$~Myr.  

In the lower half of this panel, we see that the cold initial
conditions produce a similar collapse or contraction in all
simulations resulting in small error-bars for $R_{\rm h}$.  The
steeper potential of the Plummer BG helps to damp the oscillations
rather quickly.  It settles at about the same value as in the uniform,
cold case.

In all four cases we approximately obtain the same values of MS at the
end of the simulations.  Both methods to measure MS tend to agree
within the $1\sigma$ deviations with the restricted method showing
slightly higher values.  The agreement seems to be better when a
Plummer background was used.  The simulations with uniform background
have shorter values of $t_{\rm seg}$, while we see no difference in this
value for the different initial virial states.  In all cases a
significant level of MS is reached in times shorter than $1$~Myr.

All four panels show an initial contraction or collapse of $R_{\rm h}$.
The contraction is stronger in the initially 'cold' simulations.
Oscillations of $R_{\rm h}$ are stronger in the uniform background
simulations, while the Plummer background helps to damp the
oscillations immediately.   This is partly due to the different
free-fall times of stars in a Plummer sphere, avoiding a collective
oscillation.  We see no significant difference in the final values of
$R_{\rm h}$ at the end of the simulations for the different background
potentials, i.e.\ we confirm again the findings of \citet{Smi11,Smi13}
that this is a function of the initial virial state alone.

\subsection{SEG-IN}
\label{sec:seg-in}

The sets of simulations (shown in Fig.~\ref{fig:PSEGIN}) are designed
to have a high value of MS initially, as we have forced the high-mass
stars to be in the central area.  In all panels we see initial
$\Lambda$ values in the range between $5$ and $6$.  Also both methods
agree at the beginning perfectly as not a single high-mass star is
ejected yet.  In all simulations we see a rapid decline of the MS
values until they settle at roughly the same values as in the NO-SEG
cases.  

This is in contrary to any prediction of the dynamics of MS in
which we should always see an increase.  The initial contraction
or even collapse (cold simulations) compacts the low-mass stars
towards the potential centre.  We indentify this initial contraction
of the low-mass stars as the main reason for the fastly declining MS
values.  In all four panels we see an initial decline of $r_{8}$ as
well, i.e.\ also the already centrally concentrated high-mass stars
are moving towards the centre as well.  This should increase the level
of MS; instead we see a rapid decrease in both methods.  While in the
non-restricted method this could be due to ejected high-mass stars,
our restricted method is designed to avoid such results.

The behaviour of $R_{\rm h}$ is similar to the NO-SEG cases explained
above. 

The top-left panel again shows the uniform BG--$Q_{\rm init} = 0.5$
case.  The simulations reach a final value of $\Lambda_{\rm final} =
2.69 \pm 0.69$ with $t_{\rm seg}={\rm always}$ (in fact all
panels show $t_{\rm seg}={\rm always}$).  The top-right panel shows
the results for the uniform BG potential starting in a cool state.
Here the final value is $2.56 \pm 0.66$.  For the Plummer BG
with $Q_{\rm init} = 0.5$ (bottom left panel) we get $\Lambda_{\rm
  final} = 2.89 \pm 0.82$, which is slightly higher than with
the uniform BG.  It seems that a steeper potential, even though not
influencing the final value of $R_{\rm h}$ is able to retain the
massive stars in the centre better than the uniform BG.
Finally, in the bottom right (Plummer BG--$Q_{\rm   init} = 0.2$) we
get $\Lambda_{\rm final} = 2.79 \pm 0.68$. 

\begin{table*}
\centering
\caption{Summary of results. The first, second and third column are
  giving information of the initial conditions: type of BG potential,
  mass distribution and virial ratio respectively.  The fourth and
  fifth column show the final values for $\Lambda_{\rm MSR}$ and
  $\Lambda_{\rm rest}$ respectively. The values are calculated by
  taking an average value of all measurements between 4 and 5~Myr.
  The sixth column gives $t_{\rm seg}$ i.e.\ the information when
  ($t_{\rm seg}$) $\Lambda_{\rm rest}$ stays larger than 2, i.e., a
  significant level of MS is detected. The last two columns are the
  final values for $r_8$ and $R_h$ respectively. Again we calculate
  them as averages from all measurements between 4 and 5~Myr.} 
\label{tab:results}
\begin{tabular}{ccccccccc}
\hline
Potential & Mass placement & $Q_{\rm init}$ & $\Lambda_{\rm MSR, final}$  &
$\Lambda_{\rm final}$ & $t_{\rm seg}$ [Myr] & $r_8$ [pc]& $R_h$ [pc]   \\
\hline \hline
Uniform & NO-SEG & $0.5$&$2.35\pm 0.94$&$2.68\pm 0.86$&$0.4$&$0.82\pm 0.20$&$0.79\pm 0.14$\\
Uniform & NO-SEG & $0.2$&$2.32\pm 1.38$&$2.76\pm 0.82$&$0.4$&$0.69\pm 0.19 $&$0.65\pm 0.11$\\
Plummer & NO-SEG & $0.5$&$2.65\pm 1.53$&$2.51\pm 0.78$&$0.8$&$0.90\pm 0.24$&$0.86\pm 0.17$\\
Plummer & NO-SEG & $0.2$&$2.78\pm 1.13$&$2.74\pm 0.74$&$0.6$&$0.65\pm 0.12$&$0.64\pm 0.10$\\
\hline
Uniform & SEG-IN & $0.5$&$2.58\pm 1.24$&$2.69\pm 0.69$&always&$0.77\pm 0.17$&$0.73\pm 0.13$\\
Uniform & SEG-IN & $0.2$&$2.03\pm 1.39$&$2.56\pm 0.66$&always&$0.71\pm 0.21$&$0.64\pm 0.11$\\
Plummer & SEG-IN & $0.5$&$3.49\pm 2.28$&$2.89\pm 0.82$&always&$0.79\pm 0.16$&$0.77\pm 0.14$\\
Plummer & SEG-IN & $0.2$&$2.20\pm 1.55$&$2.79\pm 0.68$&always&$0.64\pm 0.18$&$0.61\pm 0.12$\\
\hline
Uniform & SEG-OUT & $0.5$&$1.43\pm 0.58$&$2.12\pm 0.57$&3.5&$1.02\pm 0.32$&$0.87\pm 0.14$\\
Uniform & SEG-OUT & $0.2$&$1.72\pm 0.91$&$2.70\pm 0.72$&1.3&$0.68\pm 0.16$&$0.63\pm 0.07$\\
Plummer & SEG-OUT & $0.5$&$1.18\pm 0.50$&$1.61\pm 0.39$&never&$1.22\pm 0.31$&$0.97\pm 0.16$\\
Plummer & SEG-OUT & $0.2$&$1.43\pm 0.69$&$2.14\pm 0.60$&3.8&$0.86\pm 0.38$&$0.69\pm 0.12$\\
\hline \hline
\end{tabular}
\end{table*}

\subsection{SEG-OUT}
\label{sec:sec-out}

In this section we describe our results (shown in
Fig.~\ref{fig:PSEGOUT}) obtained with rather unusual initial
conditions \citep[even though some authors claim to measure inverse
mass segregation e.g.\ in Taurus][]{Par11}, i.e.\ we start with all
high-mass stars being formed in the outskirts ($> 1$~pc) of the
stellar distribution.  This is inverse MS and should be reflected by
initial $\Lambda$ values below unity.  Still, as a mean value taken
from $40$ simulations we see a value close to unity, i.e.\ no
segregation.  While most simulations in our sample indeed have values
below unity, this rather high mean value is due to some special cases
in which many high-mass stars are located close to each other but
still outside the central area.  It is in fact a curiosity produced
due to the fractal initial distributions we are using together with
the single criterion that high-mass stars have to be outside the
central $1$~pc area.  We do not force the $\Lambda$ values to be below
unity.  In one particular simulation we even obtain a value of
$\Lambda_{\rm rest} = 1.8$. 

In the pseudo-equilibrium cases (left panels), we see no fast
evolution in MS but a slow increase with the uniform BG 
($\Lambda_{\rm final} = 2.12 \pm 0.57$ and $t_{\rm seg}=3.5$~Myr)
or almost no evolution at all with the Plummer BG ($\Lambda_{\rm
  final} =1.61 \pm 0.39$ and $t_{\rm seg} = {\rm never}$).  This shows
that the small contraction of our stellar distribution is not enough
to ensure sufficient interactions between the high-mass stars and the
lower mass stars to drive them efficiently towards the centre.  How
realistic such initial conditions are, is debatable.

If we look at the cold cases (right panels), we see again a very fast
evolution of MS.  Here, it seems the initial collapse of the stellar
distribution is enough to drag the high-mass stars to the central area
and then they stay there.  We get for the uniform BG case a value of 
$2.70 \pm 0.72$ with a $t_{\rm seg} = 1.3$~Myr and for the
Plummer BG case $2.14 \pm 0.60$ and $t_{\rm seg}=3.8$~Myr.

The bottom halves of the panels show again the same evolution of
$R_{\rm h}$ as seen in the other sets, but this time we note that the
values for $r_{8}$ are always larger than $R_{\rm h}$ for all initial
conditions and throughout the simulation times.  This means that in
all simulations we never got 8 or more massive stars within the
half-mass radius.
 
\section{Summary \& Discussion}
\label{sec:discussion}

A summary of the most important initial parameters and the associated
results is given in Tab.~\ref{tab:results}.  

In Fig.~\ref{fig:massdistfin} we show the final configurations for the 
three example simulations shown in Fig.~\ref{fig:massdist}.  In the
left and central panel one clearly sees the effect of MS.  In the
right panel (SEG-OUT case) the results are debatable, but still an
over-density of massive stars can be detected in the centre.  In fact
this particular example has indeed a final value higher than two.

All sets of simulations (except the SEG-OUT cases) reach values of
$\Lambda_{\rm rest} \geq 2$ very fast (less than 1~Myr) or even show
these values from the start (SEG-IN).

This shows that it is possible to observe MS after very short times in
the embedded phase for clusters which are not initially segregated, so
the premise where MS has to be primordial
\citep[e.g. ][]{Bon&Dav98,Hil&Har98,Rab&Mer98} is not strictly
necessary.  \cite{Par15} found that primordial MS is less common when
in the process of star formation the influence of feed-back is taken
into account.  However, we cannot exclude primordial MS, which is
supported by hydrodynamical simulations
\citep[e.g. ][]{Kle01b,Bon01a,Bon&Bat06} or even a mix of both effects
as shown by \citet{Kuz15}.

After gas-expulsion, the naked SC can inherit the level of MS from the
embedded phase, which can be used as an initial condition for studies
of later evolution \citep[e.g. ][]{Mcm07,Moe&Bon09b,All09}.

The global evolution, in the scenario SEG-OUT, is slower
due the location of the massive stars.  The low gravitational
potential of a Plummer BG in the outer parts joined with a weak
compression ($Q_{\rm init} = 0.5$) is not producing a significant
level of MS.  In comparison, the flat uniform BG, which is stronger in 
the outer parts, is helping to drag most of the massive stars to the
centre.  For the cool initial state, the Plummer BG is helping the
massive stars to reach locations with a deeper potential but not
within a short time interval.  

\citet{All10}, \citet{Yu11} and \citet{Par14} performed similar N-body
simulations but without any background potential for the gas
distribution.  When starting with initially virial (in reality
pseudo-virial) conditions they obtain MS values around $\sim 2$.  In
their cold simulations they obtain MS values of about $6$
\citep{All10} or even larger than $20$ \citep{Yu11}.  While our MS
values in the pseudo-virial case are somewhat higher they mainly agree
within the errors.  In fact we obtain similar results as the median
values of \citet{Par14}.  We believe the reason for the differences
can be found in the use of the background potential which is helping
to retain stars (in this case high-mass stars) in the central cluster
area, which otherwise would have been ejected or at least transported
to larger radii.  In the initially cold simulations, we see a strong
collapse of all stars alike, transporting many low-mass stars to the
central areas as well.  This lowers our values of MS significantly and
by using a background potential, we also keep more low-mass stars from
being ejected.  

We have to discuss the different values of the $\Lambda_{\rm MSR}$ and
$\Lambda_{\rm rest}$ parameters in our simulations.  We suggest that
the use of $\Lambda_{\rm rest}$ produces more reliable results, even
though due to the radius restriction for the low-mass stars, the
values are slightly higher than in the unrestricted case.  We have
tested different radius criterions for the low-mass stars.  If we
would use the same radius as for the high-mass stars we would always
get $\Lambda_{\rm rest}$ values close to unity.  Without restriction
we would have very similar values than the unrestricted case
($\Lambda_{\rm MSR}$).  Therefore our criterion of a restrcition
radius of double the length of $r_{8}$ or $R_{\rm h}$ respectively
seems to give the most reliable results, without dealing with escapers
or high-mass stars orbiting the outskirts of a cluster.  This also
leads to smaller error bars than in the unrestricted method.

Running our simulations leads to a contraction of the distribution in
the pseudo-virialised case or to a collapse in the cold simulations,
resulting in a more spherical symmetric and far more denser
distribution of the stars.  This process is very fast (less than
1~Myr) and very effective in depositing almost all high-mass stars in
the central area (NO-SEG and SEG-IN cases; see left and middle panel
in Fig.~\ref{fig:massdistfin}).  Even in the SEG-OUT case we do see 
many high-mass stars in the centre.  Per definition this should be a
highly mass-segregated configuration.  But not only the high-mass
stars are now centrally concentrated, the low-mass stars are also more
centrally concentrated than at the beginning, leading to a high
probability for short MSTs, which enter the calculation of
$\Lambda_{\rm MSR}$ and even more so for $\Lambda_{\rm rest}$.   This
will lead to lower values of MS, and explains why in the SEG-IN case
we see a sharp initial drop of the $\Lambda$ values, due to all the
low-mass stars settling into a more concentrated distribution.

In our models, no stellar evolution was included which is not
realistic, but as we reach significant levels of MS in less than one
Myr, i.e.\ a time in which even the most massive stars of our
simulations do not evolve, we are confident that this short-fall of
our simulations would not alter the final results.  The influence of
stellar evolution on our results will be subject of a different
investigation in the near future.

\section{Conclusions}
\label{sec:conc}

We have presented simulations of the early evolution of initially cool
or pseudo-virial, fractal embedded star clusters.  The clusters start
with three different initial levels of MS namely NO-SEG, SEG-IN and
SEG-OUT.  We follow the level of MS using the MST method using all the
stars ($\Lambda_{\rm MSR}$) and the same method restricted to $R_h$ or
$r_{8}$ ($\Lambda_{\rm rest}$). 

We conclude that in almost all cases (some exceptions with the SEG-OUT
conditions), due to the contraction or collapse (pseudo-virial or cold
conditions, respectively) we get significant MS within a short
time-span of less than 1~Myr.  Or, if one refuses to call our initial
conditions a star cluster (i.e., if the term star cluster is only
applied to a somewhat spherical and centrally concentrated
over-density of stars), then our results imply that as soon as {\it
  the} star cluster is formed it is automatically mass-segregated from
the start. 

On the other hand our results show, that the distribution of stars
inside the cluster have lost any information of their initial birth
place and so the question if stars are born mass-segregated cannot be
answered by looking at a mass-segregated cluster.  The only set of
simulations which retained at least part of this initial information
is the SEG-OUT case, i.e.\ the most massive stars are
born in the outer parts of the star forming region.  Here, some
massive stars on rather circular orbits in the outskirts of the
cluster (star forming region) do not fully take part in the
initial contraction phase and therefore do not get deposited into the
central area.

To answer the question if the MS is independent of the properties of
the gas in embedded young star clusters, we find a similar degree of
MS ($\Lambda_{\rm final}$) independent of the conditions of the gas we
use. 

We conclude that any combination of initial virial ratio, BG
potential, and initial mass distribution can lead to almost the same
levels of MS with $\Lambda_{\rm final} \approx 2.6\pm^{0.8}_{0.5}$.  
The only exception is the SEG-OUT scenario together with
pseudo-virial velocities, where this parameter shows somewhat lower
values.   

The dependence on the stochastic initial conditions of the fractal
distributions produces in a few cases results far away from the
average values reported in this study and we cannot exclude the
possibility that these cases can be observed in reality or in
theoretical studies, which may be much more sophisticated (and
therefore more time consuming) but have to rely on one or a few
simulations only. \\ 

{\bf Acknowledgments:}
MF acknowledges financial support of FONDECYT grant No.~1130521,
Conicyt PII20150171 and BASAL PFB-06/2007.  RD is partly funded
through a studentship of FONDECYT grant No.~1130521 and BASAL
PFB-06/2007.  

\bibliographystyle{mnras}

\label{lastpage}

\end{document}